\documentclass[journal]{IEEEtran}
\usepackage[pdftex]{graphicx}

\usepackage{amsmath,amssymb,amsfonts,stmaryrd}
\usepackage{bm}
\usepackage{dsfont}
\usepackage{numprint}
\newcommand{\isep}{\mathrel{{.}\,{.}}\nobreak}

\usepackage{amsthm}

\usepackage{adjustbox}
\usepackage{booktabs}
\usepackage{multirow}

\usepackage[caption=false,font=footnotesize]{subfig}
\usepackage{tikz}
\usetikzlibrary{matrix, positioning, patterns, shapes, arrows}
\usepackage{pgfplots}
\pgfplotsset{compat=newest}
\usepgfplotslibrary{groupplots}
\usepgfplotslibrary{colorbrewer}
\usetikzlibrary{spy,backgrounds}

\usepackage{lipsum}
\usepackage{url}

\usepackage{fancyhdr}

\definecolor{Paired-1}{RGB}{31,120,180}
\definecolor{Paired-2}{RGB}{166,206,227}
\definecolor{Paired-3}{RGB}{51,160,44}
\definecolor{Paired-4}{RGB}{178,223,138}
\definecolor{Paired-5}{RGB}{227,26,28}
\definecolor{Paired-6}{RGB}{251,154,153}
\definecolor{Paired-7}{RGB}{255,127,0}
\definecolor{Paired-8}{RGB}{253,191,111}
\definecolor{Paired-9}{RGB}{106,61,154}
\definecolor{Paired-10}{RGB}{202,178,214}
\definecolor{Paired-11}{RGB}{177,89,40}
\definecolor{Paired-12}{RGB}{105,105,105}
\definecolor{Paired-13}{RGB}{80,80,80}

\usepackage[ruled, linesnumbered, vlined]{algorithm2e}

\newcommand{\furkan}[2]{\ifx&#2&{\leavevmode\color{magenta}#1}\else{\leavevmode\color{magenta}FURKAN\{}#1{\leavevmode\color{magenta}\}}\footnote{{\leavevmode\color{magenta}#2}}\PackageWarning{Furkan}{#1: #2}\fi}

\def\addlegendimage{\csname pgfplots@addlegendimage\endcsname}

\usepackage[draft]{fixme} 
\fxsetup{theme=color}

\DeclareMathOperator*{\argmax}{arg\,max}

 \usepackage{microtype}

\usepackage[cmintegrals]{newtxmath}

\AtBeginDocument{\setlength\abovedisplayskip{4pt}}
\AtBeginDocument{\setlength\belowdisplayskip{4pt}}

\ifCLASSINFOpdf
\else
\fi

\fancypagestyle{FirstPage}{

\lfoot{ Citation information: DOI 10.1109/TVLSI.2022.3223692 \copyright2022 IEEE}

}

\begin{document}
%
\title{{List-GRAND: A practical way to achieve Maximum Likelihood Decoding}}
%
%
%

\author{Syed~Mohsin~Abbas,~Marwan~Jalaleddine~and~Warren~J.~Gross,~\IEEEmembership{Senior Member,~IEEE}
\thanks{S. M. Abbas, M. Jalaleddine and W. J. Gross are with the Department of
Electrical and Computer Engineering, McGill University, Montr{\'e}al, Qu{\'e}bec,
Canada.~(email:~syed.abbas@mail.mcgill.ca,~marwan.jalaleddine@mail.mcgill.ca,~warren.gross@mcgill.ca.).
}}

\maketitle

\begin{abstract}

Guessing Random Additive Noise Decoding (GRAND) is a recently proposed universal Maximum Likelihood (ML) decoder for short-length and high-rate linear block-codes. Soft-GRAND (SGRAND) is a prominent soft-input GRAND variant, outperforming the other GRAND variants in decoding performance; nevertheless, SGRAND is not suitable for parallel hardware implementation. Ordered Reliability Bits-GRAND (ORBGRAND) is another soft-input GRAND variant that is suitable for parallel hardware implementation, however it has lower decoding performance than SGRAND. In this paper, we propose List-GRAND (LGRAND), a technique for enhancing the decoding performance of ORBGRAND to match the ML decoding performance of SGRAND. Numerical simulation results show that LGRAND enhances ORBGRAND's decoding performance by $0.5-0.75$ dB for channel-codes of various classes at a target FER of $10^{-7}$. For linear block codes of length $127/128$ and different code-rates, LGRAND's VLSI implementation can achieve an average information throughput of $47.27-51.36$ Gbps. In comparison to ORBGRAND's VLSI implementation, the proposed LGRAND hardware has a $4.84\%$ area overhead. 
\end{abstract}


%
\IEEEpeerreviewmaketitle

\section{Introduction}
\thispagestyle{FirstPage}
\IEEEPARstart{U}{RLLC} (Ultra Reliable Low Latency Communication) is considered an important use case of 5G and future communication networks because it enables applications that require high reliability and very low latency. Some of these emerging applications include augmented and virtual reality \cite{URLLC2}, Intelligent Transportation Systems (ITS) \cite{URLLC1}, the Internet of Things (IoT) \cite{IoT1,IoT2}, and Machine-to-Machine communication (M2M) \cite{URLLC3}. These novel applications benefit from the use of short-length, high-rate error-correcting codes. Guessing Random Additive Noise Decoding (GRAND) \cite{Duffy19TIT} is a recently proposed universal maximum likelihood (ML) decoding technique for these short-length and high-rate linear block codes. GRAND is a noise-centric and code-agnostic decoder, which implies that, unlike traditional decoding techniques, GRAND attempts to guess the noise that corrupted the codeword during transmission through the communication channel. {Therefore, GRAND can be used with both structured codes and unstructured codes, which are stored in a dictionary, provided that there exists a method to verify codebook membership of a given vector \cite{DuffySRGRAND}}. Furthermore, when used with random-codebooks, GRAND achieves capacity \cite{Duffy19TIT}. 
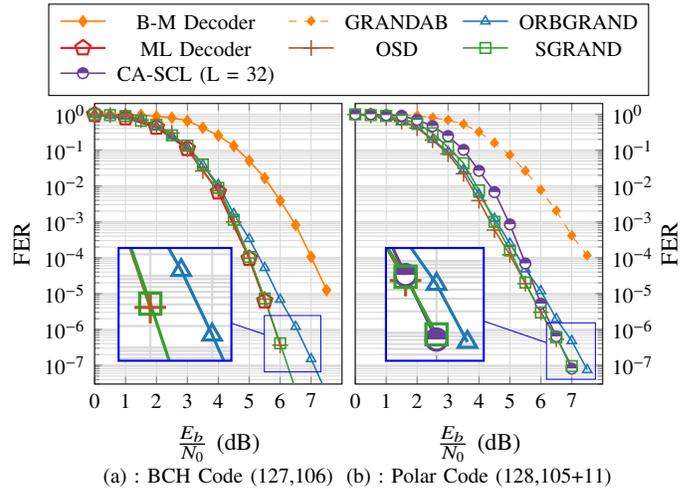
\begin{figure}[!t]
  \centering
  \begin{tikzpicture}[spy using outlines = {rectangle, magnification=2.0, connect spies}]
    \begin{groupplot}[group style={group name=fer_queries, group size= 2 by 1, horizontal sep=5pt, vertical sep=5pt},
      footnotesize,
      height=.6\columnwidth,  width=0.55\columnwidth,
      xlabel=$\frac{E_b}{N_0}$ (dB),
      xmin=0, xmax=8, xtick={0,1,...,7},
      ymode=log,
      tick align=inside,
      grid=both, grid style={gray!30},
      /pgfplots/table/ignore chars={|},
      ]

      \nextgroupplot[ylabel= FER, ytick pos=left, y label style={at={(axis description cs:-0.225,.5)},anchor=south},ymin=3e-8, ymax = 2]

     \addplot[mark=diamond*       , Paired-7 , semithick]  table[x=Eb/N0, y=FER] {data/BCH/127_106/BCH_N127_106_BM.txt};\label{gp:plot1_bch}
        \addplot[mark=triangle  , Paired-1 , semithick]  table[x=Eb/N0, y=FER] {data/BCH/127_106/BCH_N127_106_ORBGRAND.txt}; \label{gp:plot2_bch}
         \addplot[mark=pentagon,mark options={scale=1.5}       , Paired-5 , thick]  table[x=Eb/N0, y=FER] {data/BCH/127_106/BCH_N127_106_ML.txt}; \label{gp:plot5_bch}
         \addplot[mark= +,mark options={scale=1.5}      , Paired-11 , semithick]  table[x=Eb/N0, y=FER] {data/BCH/127_106/BCH_N127_106_OSD.txt}; \label{gp:plot6_bch}
         \addplot[mark=square  , Paired-3  , semithick]  table[x=Eb/N0, y=FER] {data/BCH/127_106/BCH_N127_106_SGRAND.txt}; \label{gp:plot8_bch}

      \coordinate (top) at (rel axis cs:0,1);

      \coordinate (spypoint1) at (axis cs:6.4,0.4e-6);
      \coordinate (magnifyglass1) at (axis cs:2.6,0.5e-5);

      \coordinate (spypoint2) at (axis cs:15.4,0.25e-6);
      \coordinate (magnifyglass2) at (axis cs:11.0,0.5e-5);

      \nextgroupplot[ylabel= FER, ytick pos=right,y label style={at={(axis description cs:1.340,.5)},anchor=south},ymin=3e-8, ymax = 2]
        \addplot[mark=diamond*       , Paired-7 , dashed]  table[x=Eb/N0, y=FER] {data/Polar/128_105/POLAR_N128_105_GRANDAB.txt};\label{gp:plot1_polar}
        \addplot[mark=triangle  , Paired-1 , semithick]  table[x=Eb/N0, y=FER] {data/Polar/128_105/POLAR_N128_105_ORBGRAND.txt}; \label{gp:plot2_polar}
         \addplot[mark= +,mark options={scale=1.5}      , Paired-11 , semithick]  table[x=Eb/N0, y=FER] {data/Polar/128_105/POLAR_N128_105_OSD.txt}; \label{gp:plot6_polar}
         \addplot[mark=halfcircle*  , Paired-9, semithick]  table[x=Eb/N0, y=FER] {data/Polar/128_105/POLAR_N128_105_L_32.txt}; \label{gp:plot7_polar}
         \addplot[mark=square  , Paired-3 , semithick]  table[x=Eb/N0, y=FER] {data/Polar/128_105/POLAR_N128_105_SGRAND.txt}; \label{gp:plot8_polar}

      \coordinate (bot) at (rel axis cs:1,0);
    \end{groupplot}
    \node[below = 1cm of fer_queries c1r1.south] {\footnotesize (a) : BCH Code (127,106)};
    \node[below = 1cm of fer_queries c2r1.south] {\footnotesize (b) : Polar Code (128,105+11)};
    \path (top|-current bounding box.north) -- coordinate(legendpos) (bot|-current bounding box.north);
    \matrix[
    matrix of nodes,
    anchor=south,
    draw,
    inner sep=0.2em,
    draw
    ]at(legendpos)
    {
      \ref{gp:plot1_bch}& \footnotesize B-M Decoder &[1pt]
      \ref{gp:plot1_polar}& \footnotesize GRANDAB  &[1pt]
      \ref{gp:plot2_bch}& \footnotesize ORBGRAND  \\ 
      \ref{gp:plot5_bch}& \footnotesize ML Decoder &[1pt]
      \ref{gp:plot6_bch}& \footnotesize OSD &[1pt]
      \ref{gp:plot8_bch}& \footnotesize SGRAND \\
      \ref{gp:plot7_polar}& \footnotesize CA-SCL (L = 32) \\
      }; 
        
      \spy [blue, width=1.5cm, height=1.5cm] on (spypoint1) in node[fill=white] at (magnifyglass1);
      \spy [blue, width=1.3cm, height=1.5cm] on (spypoint2) in node[fill=white] at (magnifyglass2);
  \end{tikzpicture}
  \vspace*{-2em}
  \caption{\label{fig:fer_bch_polar_ini} {Comparison of the decoding performance of different GRAND variants for BCH Code (127,106) and 5G-NR Polar Code (128,105+11).}}
  \vspace*{-1em}
\end{figure}

GRAND and its variants work on the premise of guessing the channel-induced noise by first generating test error patterns ($\bm{e}$), then applying them to the received hard-demodulated vector of channel observation values ($\hat{\bm{y}}$), and finally querying the resulting vector ($\hat{\bm{y}}\oplus\bm{e}$) for codebook membership. The order in which these test error patterns are generated is the primary difference between the GRAND variants. GRAND with ABandonment (GRANDAB) \cite{Duffy20205g},\cite{Duffy19TIT} is a hard decision input variant that generates test error patterns in ascending Hamming weight order, up to the weight $AB$. Ordered Reliability Bits GRAND (ORBGRAND) \cite{duffy2020ordered} and Soft GRAND (SGRAND) \cite{solomon2020soft} are soft-input variants that efficiently leverage soft information (channel observation values ($\bm{y}$)), resulting in improved decoding performance compared to the decoding performance of the hard-input GRANDAB. In comparison to other code-agnostic channel code decoders such as brute-force ML decoding and Ordered Statistic Decoding (OSD) \cite{Fossorier95}-\cite{OSD5}, GRAND offers a low complexity decoding solution for short-length and high-rate channel codes. The scope of this work is restricted to short channel codes with high code rates because GRAND and its variants are proposed as ML decoders for short-length and high-rate channel codes. 

Figure \ref{fig:fer_bch_polar_ini} (a) compares the decoding performance of different variants of GRAND with Berlekamp-Massey (B-M) \cite{Berlekamp68,Massey69} decoder, OSD ($\text{Order}=2$) and ML decoding of BCH code $(127,106)$. The ML decoding results are obtained from \cite{kaiserslautern}. The numerical simulation results presented in this work are based on BPSK modulation over an Additive White Gaussian Noise (AWGN) channel. While both soft-input variants of GRAND (ORBGRAND and SGRAND) outperform the hard-input B-M decoder, SGRAND achieves ML performance similar to OSD. Figure \ref{fig:fer_bch_polar_ini} (b) compares the decoding performance of various GRAND variants for decoding 5G New Radio (NR) CRC-aided polar code (128,105+11).  Furthermore, the decoding performance of state-of-the-art soft-input decoders such as the CRC-aided Successive Cancellation List (CA-SCL) decoder \cite{Tal15,LLR-List} and OSD ($\text{Order}=2$) is included for reference. The ORBGRAND and SGRAND outperform the hard-input GRANDAB ($AB=3$) variant in decoding performance, and the SGRAND achieves ML decoding performance similar to OSD, as shown in Fig. \ref{fig:fer_bch_polar_ini} (b).

As shown in Fig. \ref{fig:fer_bch_polar_ini}, SGRAND outperforms the other GRAND variants in terms of decoding performance; however, SGRAND is not suitable for parallel hardware implementation. The test error patterns generated by SGRAND are interdependent, and their query order varies with each received vector of channel observation values ($\bm{y}$) (explained in section \ref{sec:TEPGen}). As a result, SGRAND does not lend itself to efficient parallel hardware implementation, and a sequential hardware implementation will result in high decoding latency, rendering it unsuitable for applications which require ultra-low latency. The ORBGRAND, on the other hand, generates test error patterns in a predetermined logistic weight order based on integer partitioning. The test error patterns generated are mutually independent and can be generated in parallel. ORBGRAND is thus highly parallelizable and well suited to parallel hardware implementation. In \cite{ORBGRAND-TVLSI}, a VLSI architecture for ORBGRAND for $n=128$ is presented, which can perform $1.16\times10^{5}$ codebook membership queries in $\numprint{4226}$ clock-cycles due to parallel generation of test error patterns in hardware. 

{In this paper, we propose List-GRAND (LGRAND), a technique for boosting the decoding performance of ORBGRAND in order to achieve ML decoding performance comparable to SGRAND. The idea behind the proposed LGRAND is to generate a list during the decoding process and choose the candidate with the highest likelihood to be the final one. The proposed LGRAND technique is not limited to the ORBGRAND test error pattern generation; rather, it can be used with any GRAND variant that uses a suboptimal test error patterns generation scheme. However, since the ORBGRAND test error pattern generation is hardware friendly and the ORBGRAND archives good error decoding performance as a soft-input decoder, we use the ORBGRAND as a baseline to present our proposed LGRAND technique to achieve the decoding performance similar to a ML decoder such as SGRAND.}

The proposed LGRAND introduces parameters that can be adjusted to match the target decoding performance and complexity budget of a specific application. For channel codes of different classes (Bose–Chaudhuri–Hocquenghem (BCH) codes \cite{Hocquenghem59,Bose1960}, Cyclic Redundancy Check (CRC) codes \cite{Peterson61} and CRC-Aided-Polar (CA-Polar) codes \cite{Arikan09}), the proposed LGRAND achieves decoding performance similar to SGRAND. LGRAND also achieves a $0.5-0.75$dB performance gain over ORBGRAND at a target FER of $10^{-7}$. Furthermore, because the proposed LGRAND algorithm is based on ORBGRAND, LGRAND lends itself well to parallel hardware implementation. The VLSI implementation results show that the proposed LGRAND can achieve an average information throughput of $47.27-51.36$ Gbps for linear block codes of length $127/128$ and different code-rates. In comparison to the ORBGRAND hardware, the proposed LGRAND hardware has a $4.84\%$ area overhead. Furthermore, as long as the length and rate constraints are met, the proposed LGRAND hardware can be used to decode any code. 

The rest of this work is structured as follows: Section II contains preliminary information on GRAND and ORBGRAND. Section III discusses the generation of test error patterns as well as the computational complexity of GRAND and its variants. Section IV presents the proposed List-GRAND (LGRAND) technique, which is used to improve the decoding performance of ORBGRAND. The numerical simulation results are presented in Section V. Section VI describes the proposed LGRAND hardware architecture as well as the implementation results. Finally, in Section VII, concluding remarks are presented.

\section{Preliminaries}
\subsection{Notations}
Matrices are denoted by a bold upper-case letter ($\bm{M}$), while vectors are denoted with bold lower-case letters ($\bm{v}$). The transpose operator is represented by $^\top$. The number of $k$-combinations from a given set of $n$ elements is noted by $\binom{n}{k}$. $\mathds{1}_n$ is the indicator vector where all locations except the $n^{\text{th}}$ are $0$ and the the $n^{\text{th}}$ is $1$. All the indices start at $1$. For this work, all operations are restricted to the Galois field with 2 elements, noted $\mathbb{F}_2$. Furthermore, we restrict ourselves to $(n,k)$ linear block codes, where $n$ is the code length and $k$ is the code dimension.
\begin{algorithm}[t]
\caption{\label{alg:ORBgrand}ORBGRAND Algorithm}
    \DontPrintSemicolon
    \SetAlgoVlined  
    \SetKwData{e}{$\bm{e}$}
    \SetKwData{p}{$\bm{p}$}
    \SetKwData{s}{$\bm{S}$}
    \SetKwData{ind}{$\bm{ind}$}
    \SetKwData{LLR}{$\bm{y}$}
    \SetKwData{sortSet}{$[\bm{r},\bm{ind}]$}
    \SetKwData{estm}{$\hat{\bm{u}}$}
    \SetKwData{ginv}{$\bm{G}^{-1}$}
    \SetKwData{LW}{${LW_\text{max}}$}
    \SetKwData{HW}{${HW_\text{max}}$}
    \SetKwData{yhat}{$\hat{\bm{y}}$}
    \KwIn{\LLR, $\bm{H}$, \ginv, \LW, \HW}
    \KwOut{\estm}
    \SetKwFunction{RecursiveComputeLLRs}{recursiveComputeLLRs}
    \SetKwFunction{HammingWeight}{HammingWeight}
    \SetKwFunction{RDecodeRONE}{redecodeR1}
    \SetKwFunction{DecodeRZERO}{decodeR0}
    \SetKwFunction{Find}{findCandidate}
    \SetKwFunction{new}{generateErrorPattern}
    \SetKwFunction{intPartition}{generateAllIntPartitions}
    \SetKwFunction{Sort}{sortChannelObservationValues}
\eIf{$\bm{H} \cdot\yhat^\top == \bm{0}$}
{\KwRet{$\estm \leftarrow\yhat\cdot\ginv$}}
{
    $\ind \leftarrow$ \Sort{\LLR} \tcp*[r]{$\lvert{\bm{y}}_i\rvert\leq\lvert{\bm{y}}_j\rvert~~\forall i < j$}
    $\e \leftarrow \bm{0}$\;
    \For{$i \gets 1$ to \LW}{
        $\s \leftarrow$ \intPartition{i}\tcp*[r]{$(\lambda_1, \lambda_2, \ldots, \lambda_P) \vdash i$  $,~\forall P \in [1, HW_\text{max}]$}\;
        \ForAll{$\bm{l}$ in \s}{
          $\e \leftarrow$ \new{$\bm{l}$,\ind}\;
          \If{$\bm{H} \cdot(\yhat \oplus \e)^\top == \bm{0}$} {
            $\estm \leftarrow (\yhat \oplus \e)\cdot\ginv$\;
            \KwRet{\estm}
            }
        }
    }
   } 
\end{algorithm}

\subsection{GRAND Decoding}

For a $(n,k)$ linear block code with codebook $\mathcal{C}$, a vector $\bm{u}$ of size $k$ maps to a vector $\bm{c}$ of size $n$, and the ratio  $R \triangleq \frac{k}{n}$ is known as the code-rate. Furthermore, there exists a $k \times n$ matrix $\bm{G}$ called generator matrix ($\bm{c}\triangleq\bm{u}\cdot\bm{G}$) and a $(n-k)\times n$ matrix $\bm{H}$ called parity check matrix.

GRAND \cite{Duffy19TIT} attempts to guess the noise that corrupted the transmitted codeword ($\bm{c}$) as it passed through the communication channel. To that end, GRAND first generates the test error patterns ($\bm{e}$) starting from the most likely up to the least likely pattern taking into account the channel model. This is followed by combining the generated test error patterns with the hard decided received vector of channel observation values (demodulated symbols) $\hat{\bm{y}}$, and evaluating if the resulting vector $\hat{\bm{y}} \oplus \bm{e}$ is a member of the codebook ($\mathcal{C}$). If the resulting vector is a member of the codebook, the decoding is assumed to be successful, and $\bm{e}$ is declared as the guessed noise, whereas $\hat{\bm{c}} \triangleq \hat{\bm{y}}~\oplus~\bm{e}$ is outputted as the estimated codeword.

GRAND can be used with any codebook as long as there is a method for validating a vector's codebook membership. For any linear codebook ($\mathcal{C}$), the codebook membership of a vector can be verified using the underlying code's parity check matrix $\bm{H}$, as follows:

\begin{equation}
\forall~\bm{c} \in \mathcal{C},~\bm{H}\cdot{\bm{c}}^\top = \bm{0}.
\label{eq:constraint}
\end{equation}

For other non-structured codebooks, stored in a dictionary, the codebook membership of a vector can be checked with a dictionary lookup. For the rest of the discussion, we restrict ourselves to $(n,k)$ linear block codes. 

\subsection{ORBGRAND Decoding}
ORBGRAND \cite{duffy2020ordered} is centered around generating distinct integer partitions of a particular logistic weight ($LW$), and these integer partitions are then used to generate test error patterns ($\bm{e}$). The logistic weight ($LW$) corresponds to the sum of the indices of non zero elements in the test error patterns \cite{duffy2020ordered}. For example, $\bm{e} = [1,1,0,0,1,0]$ has a Hamming weight of $3$, whereas the logistic weight is $1+ 2 + 5 = 8$.

An integer partition $\bm{\lambda}$ of a positive integer $m$, noted $\bm{\lambda} = (\lambda_1, \lambda_2, \ldots, \lambda_P) \vdash m$ where $\lambda_1>\lambda_2>\ldots>\lambda_P$, is the multiset of positive integers $\lambda_i$ $(\forall i \in [1, P])$ that sum to $m$. If all parts $\lambda_i$ $(\forall i \in [1, P])$ of the integer partition are different, the partition is called distinct. Please note that the Hamming weight of the generated test error pattern ($\bm{e}$) obtained from an integer partition $\bm{\lambda} = (\lambda_1, \lambda_2, \ldots, \lambda_P)$ with $P$ elements is $P$. ORBGRAND considers the maximum logistic weight for a $(n,k)$ linear block code to be $\frac{n(n+1)}{2}$ ($LW_\text{max} = \frac{n(n+1)}{2}$). Furthermore, the generated test error patterns have a maximum Hamming weight of $n$ ($HW_\text{max} = n$). It should be noted that only distinct integer partitions are considered for generating test error patterns, and all the parts ($\lambda_i$) of the integer partitions are less than or equal to $n$ ($\lambda_i\leq~n$ $(\forall i \in [1, P])$) \cite{duffy2020ordered}.

Algorithm \ref{alg:ORBgrand} summarizes the steps of the ORBGRAND. The inputs to the algorithm are the vector of channel observation values (log-likelihood ratios (LLRs)) $\bm{y}$ of size $n$, a $(n-k)\times n$ parity check matrix $\bm{H}$, a $n\times k$ matrix $\bm{G}^{-1}$ where $\bm{G}^{-1}$ refers to the inverse of the generator matrix $\bm{G}$ of the code such that $\bm{G}\cdot \bm{G}^{-1}$ is the $k\times k$ identity matrix, and the maximum Hamming weight $HW_\text{max}$ as well as the maximum logistic weight considered $LW_\text{max}$. 

The algorithm begins with evaluating the received vector's ($\hat{\bm{y}}$) codebook membership (line 1); if it is satisfied (\ref{eq:constraint}), the original message is retrieved (line 2); otherwise, $\bm{y}$ is sorted in ascending order according to the absolute values of the LLRs ($\lvert{\bm{y}}_i\rvert\leq\lvert{\bm{y}}_j\rvert~~\forall i < j$), and the relevant indices are recorded into a permutation vector $\bm{ind}$ (line 4). This is followed by generating all the integer partitions for each logistic weight (line 7). The function \textit{generateErrorPattern} generates a test error pattern ($\bm{e}$) using integer partition ($\bm{l}$), which is then ordered using the permutation vector $\bm{ind}$ (line 10). For instance, the generated error pattern, for $n=6$ with $\bm{l}=(1,2)$ and $\bm{ind}$ = $(2,6,5,4,3,1)$, will be $\bm{e}$ = $(0,1,0,0,0,1)$. The generated test error patterns are then applied sequentially to the hard decision vector ($\hat{\bm{y}}$), which is obtained from $\bm{y}$. The resulting vector ($\hat{\bm{y}}~\oplus~\bm{e}$) is then queried for codebook membership (line 11). If the codebook membership criterion (\ref{eq:constraint}) is met, then $\bm{e}$ is the guessed noise and $\hat{\bm{c}} \triangleq \hat{\bm{y}}~\oplus~\bm{e}$ is the estimated codeword. Otherwise, either the remaining error patterns for that logistic weight or larger logistic weights are considered. Finally, using $\bm{G}^{-1}$ (line 12), the original message ($\hat{\bm{u}}$) is retrieved from the estimated codeword, and the decoding process is terminated.

\begin{figure}
  \centering
  \includegraphics[width=1\linewidth]{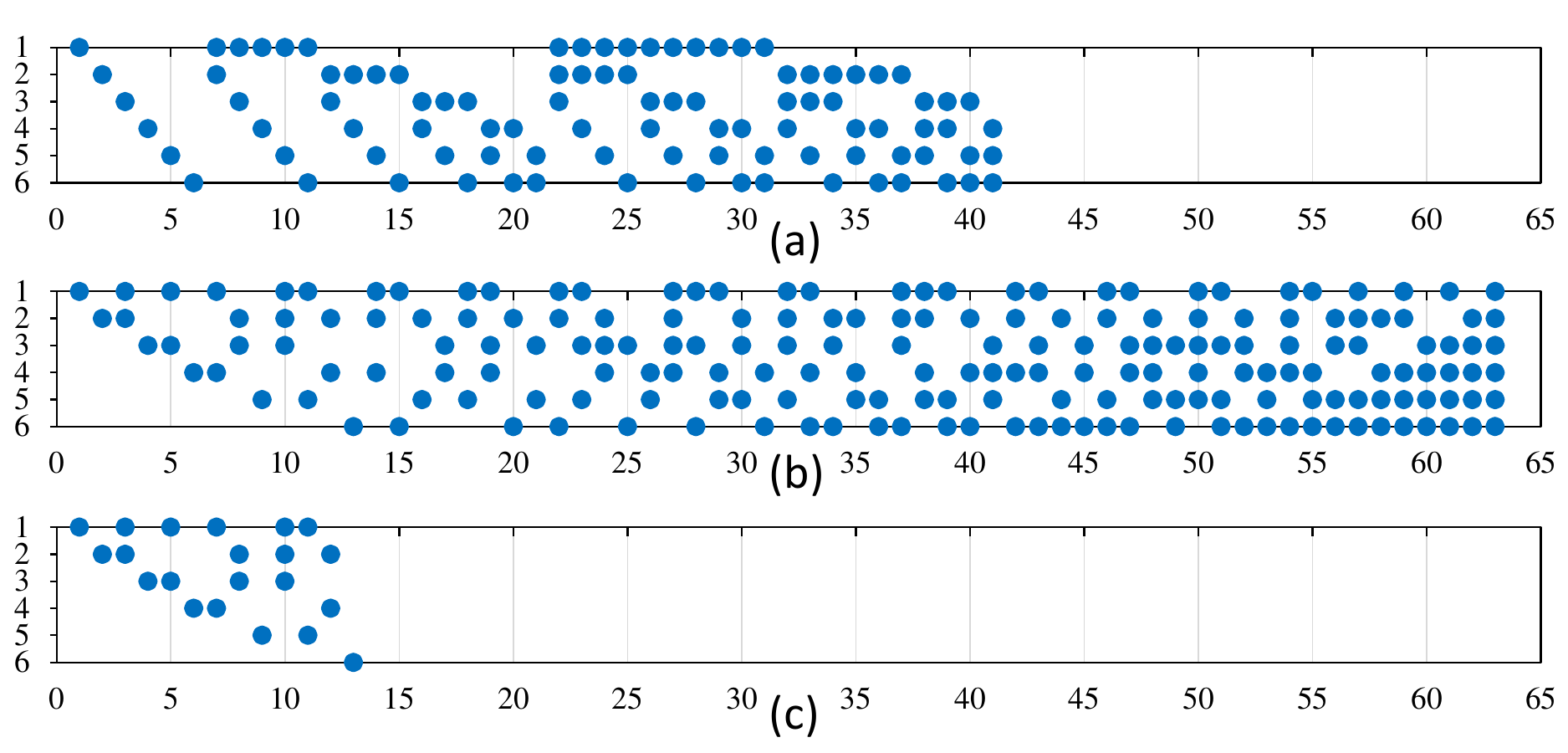}
   \vspace*{-2em}
  \caption{Test Error Pattern (TEP) generation for GRAND for  $n=6$ (a) Upper: TEP generation for GRANDAB ($AB=3$) (b) Middle: TEP generation for ORBGRAND ($LW_\text{max} = 21$) (c) Bottom: TEP generation for ORBGRAND ($LW_\text{max} = 6$)}
  \label{fig:TEP_sch} 
   \vspace*{-1em}
\end{figure}

\begin{figure}
  \centering
  \includegraphics[width=1\linewidth]{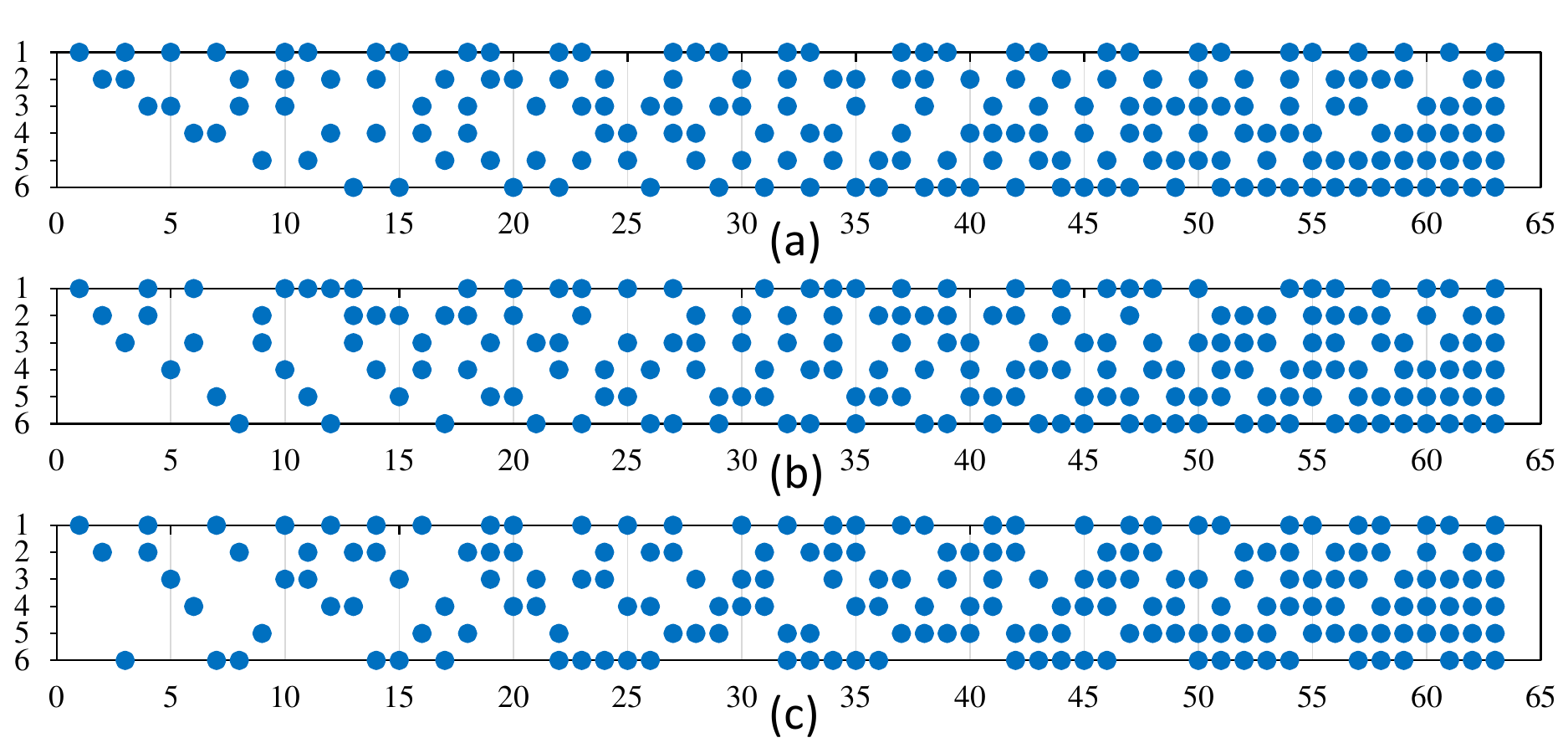}
   \vspace*{-2em}
  \caption{Test Error Pattern (TEP) generation for SGRAND (a) {Upper: ML order for $\bm{y_1}$=$[1.0,2.1,3.2,4.3,5.4,6.5]$ (b) Middle: ML order for $\bm{y_2}$=$[1.3, 2.5, 3.6, 4.9, 5.8, 6.1]$ (c) Bottom: ML order for $\bm{y_3}$=$[1.8, 2.0, 3.9, 4.1, 5.6, 3.3]$}}
  \label{fig:TEP_sch_sgrand} 
   \vspace*{-1em}
\end{figure}
\begin{figure}[!t]
  \centering
  \begin{tikzpicture}[spy using outlines = {rectangle, magnification=2.0, connect spies}]
    \begin{groupplot}[group style={group name=fer_queries, group size= 2 by 1, horizontal sep=5pt, vertical sep=5pt},
      footnotesize,
      height=.6\columnwidth,  width=0.55\columnwidth,
      xlabel=$\frac{E_b}{N_0}$ (dB),
      xmin=0, xmax=8, xtick={0,1,...,7},
      ymode=log,
      tick align=inside,
      grid=both, grid style={gray!30},
      /pgfplots/table/ignore chars={|},
      ]

      \nextgroupplot[ylabel= FER, ytick pos=left, y label style={at={(axis description cs:-0.225,.5)},anchor=south},ymin=1e-9, ymax = 2]
      \addplot[mark=diamond* , Paired-7, semithick]  table[x=Eb/N0, y=FER] {data/CRC/128_104/CRC_N128_104_GRANDAB.txt};\label{gp:plot1_c}
      \addplot[mark=triangle, Paired-1, semithick]  table[x=Eb/N0, y=FER] {data/CRC/128_104/CRC_N128_104_ORBGRAND.txt};\label{gp:plot3_c}
      \addplot[mark=square  , Paired-3, semithick]  table[x=Eb/N0, y=FER] {data/CRC/128_104/CRC_N128_104_SGRAND_10_7.txt}  ;\label{gp:plot4_c}

      \coordinate (top) at (rel axis cs:0,1);

      \coordinate (spypoint1) at (axis cs:6.7,0.5e-7);
      \coordinate (magnifyglass1) at (axis cs:2.6,0.5e-6);

      \coordinate (spypoint2) at (axis cs:15.4,0.8e-8);
      \coordinate (magnifyglass2) at (axis cs:11.0,0.8e-6);

      \nextgroupplot[ylabel=Avg. Queries, ytick pos=right,y label style={at={(axis description cs:1.325,.5)},anchor=south},ymin=1, ymax = 5e7]
      \addplot[mark=diamond* , Paired-7, semithick]  table[x=Eb/N0, y=Tests/f] {data/CRC/128_104/CRC_N128_104_GRANDAB.txt};
      \addplot[mark=triangle, Paired-1, semithick]  table[x=Eb/N0, y=Tests/f] {data/CRC/128_104/CRC_N128_104_ORBGRAND.txt};
      \addplot[mark=square  , Paired-3, semithick]  table[x=Eb/N0, y=Tests/f] {data/CRC/128_104/CRC_N128_104_SGRAND_10_7.txt}  ;

      \coordinate (bot) at (rel axis cs:1,0);
    \end{groupplot}
    \node[below = 1cm of fer_queries c1r1.south] {\footnotesize (a) : FER};
    \node[below = 1cm of fer_queries c2r1.south] {\footnotesize (b) : Avg. Queries};
    \path (top|-current bounding box.north) -- coordinate(legendpos) (bot|-current bounding box.north);
    \matrix[
    matrix of nodes,
    anchor=south,
    draw,
    inner sep=0.2em,
    draw
    ]at(legendpos)
    {
      \ref{gp:plot1_c}& \footnotesize GRANDAB (AB=3) &[1pt]
      \ref{gp:plot4_c}& \footnotesize SGRAND \\ 
      \ref{gp:plot3_c}& \footnotesize ORBGRAND ($LW_\text{max}$=8256, $HW_\text{max}$=128)  \\
       }; 
      \spy [blue, width=1.7cm, height=1.9cm] on (spypoint1) in node[fill=white] at (magnifyglass1);
  \end{tikzpicture}
  \vspace*{-2em}
  \caption{\label{fig:fer_crc_128_104}Comparison of decoding performance and average complexity GRANDAB, ORBGRAND and SGRAND {(Queries$_\text{max}$=$5\times10^{7}$)} decoding of CRC Code(128,104).}
  \vspace*{-1em}
\end{figure}
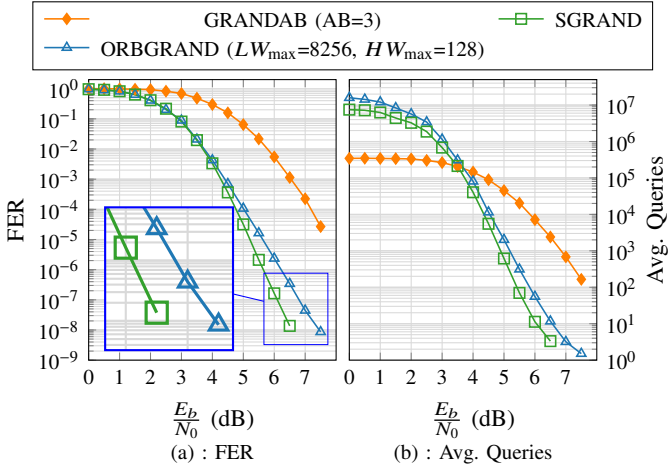

\section{GRAND: Analysis of Test Error Pattern (TEP) generation and computational complexity}

This section describes the TEP ($\bm{e}$) generating scheme and computational complexity analysis for GRAND and its variants.

\subsection{TEP generation for GRAND}\label{sec:TEPGen}
Combining a TEP with the hard demodulated received vector $\bm{\hat{y}}$ corresponds to flipping certain bits of that vector ($\bm{\hat{y}}$). GRAND with ABandonment (GRANDAB) \cite{Duffy19TIT} is a hard decision input version of GRAND that generates TEPs in increasing Hamming weight order up to a Hamming weight $AB$.  The TEPs generated in Hamming weight order for $n=6$ and $AB=3$ are depicted in Fig. \ref{fig:TEP_sch} (a), where each column corresponds to a TEP and a dot corresponds to a flipped bit location of the received hard demodulated vector ($\bm{\hat{y}}$). 

On the other hand, ORBGRAND is a soft-input GRAND variant that uses the Logistic weight order to generate TEPs. The TEPs generated by ORBGRAND with $LW_\text{max} = 21$ are shown in Fig. \ref{fig:TEP_sch} (b). The integer partitions of an integer $m$ $(\forall m \in [1, 21])$ are generated sequentially and these integer partitions are then used to generate TEPs. For $n=6$ and $LW_\text{max} = 21$,  $\numprint{63}$ TEPs are generated, with the maximum Hamming weight ($HW_\text{max}$) of the generated TEPs being $6$. However, when $LW_\text{max}$ is reduced from $21$ to $6$ the number of TEPs is reduced to $\numprint{13}$, as shown in Fig. \ref{fig:TEP_sch} (c). As a result, the parameter $LW_\text{max}$ can be adjusted to limit the maximum number of TEPs.

Soft GRAND (SGRAND) \cite{solomon2020soft} incorporates all soft information into the decoder to generate the ML order of TEPs, and an efficient method for generating the ML order can be found in \cite{valembois}. The ML order for generating TEPs for $n=6$ is shown in Fig. \ref{fig:TEP_sch_sgrand}. The ML order for TEP generation is dependent on $\bm{y}$ and changes with each new vector received (we refer the reader to Algorithm 2 of \cite{solomon2020soft} for further details about ML order TEP generation). Let $\bm{y_1}$=$[1.0,2.1,3.2,4.3,5.4,6.5]$ be the received vector of channel observation values at time instant 1, and the ML order corresponding to  $\bm{y_1}$ is shown in Fig. \ref{fig:TEP_sch_sgrand} (a). At the second time step, the received vector from the channel is $\bm{y_2}$=$[1.3, 2.5, 3.6, 4.9, 5.8, 6.1]$ and the corresponding ML order for TEP generation is depicted in Fig. \ref{fig:TEP_sch_sgrand} (b). Unlike ORBGRAND, even though the order of the absolute value of LLRs is the same for $\bm{y_1}$ and $\bm{y_2}$, the TEPs are generated in a different order for $\bm{y_1}$ and $\bm{y_2}$. Similarly, as shown in Fig. \ref{fig:TEP_sch_sgrand} (c), the ML order changes at a third time instant when $\bm{y_3}$=$[1.8, 2.0, 3.9, 4.1, 5.6, 3.3]$ changes. 

As a result of the changing TEP query order with each received vector from the channel ($\bm{y}$) and the TEP interdependence \cite{solomon2020soft}, SGRAND does not lend itself to efficient parallel hardware implementation.  Alternatively, developing a sequential hardware implementation for SGRAND will result in a high decoding latency, which is unsuitable for applications requiring ultra-low latency. ORBGRAND, on the other hand, generates TEPs in the predetermined logistic weight order. Therefore, ORBGRAND is far better suited to parallel hardware implementation than SGRAND.

\subsection{Computational Complexity of GRAND}

The computational complexity of GRAND and its variants can be expressed in terms of the number of codebook membership queries required. In GRAND and its variants, a codebook membership query consists of simple operations such as bit-flips and a syndrome check (codebook membership verification (\ref{eq:constraint})). Furthermore, the complexity can be divided into two categories: worst-case complexity, which corresponds to the maximum number of codebook membership queries required, and average complexity, which corresponds to the average number of codebook membership queries required. For a codelength of $n=128$, the worst-case number of queries for GRANDAB $(AB =3)$ decoder is $\numprint{349632}$ queries ($\sum\limits_{i=1}^{AB} \binom{n}{i}$ \cite{Duffy19TIT}). The worst-case number of queries for the ORBGRAND decoder depends on the value of the parameter $LW_\text{max}$; for example, with $LW_\text{max}=96$ and $n=128$, the worst case complexity is $3.69\times10^{6}$ queries \cite{ORBGRAND-VLSI}. For SGRAND \cite{solomon2020soft}, the parameter Queries$_\text{max}$ (Queries$_\text{max}$=$5\times10^{7}$; Fig. \ref{fig:fer_crc_128_104}), which represents the maximum number of queries allowed, determines the worst-case complexity.

Figure \ref{fig:fer_crc_128_104} compares the frame error rate (FER) performance and average complexity for different GRAND variants for decoding CRC Code (128,104). As seen in Fig. \ref{fig:fer_crc_128_104} (b), as channel conditions improve, the average complexity of GRAND and its variants decreases sharply because transmissions subject to light noise are decoded quickly \cite{Duffy19TIT}\cite{solomon2020soft}\cite{duffy2020ordered}. In terms of error decoding performance, SGRAND outperforms other GRAND variants by generating TEPs in ML order \cite{solomon2020soft},\cite{valembois}. As a result, SGRAND achieves ML decoding performance while requiring the fewest average number of codebook membership queries, as shown in Fig. \ref{fig:fer_crc_128_104}. However, as explained previously, the SGRAND is not suited for parallel hardware implementation. 


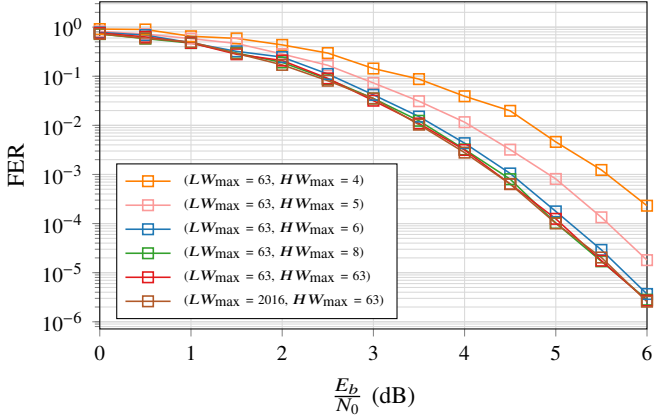
\begin{figure}[!t]
 \centering
\begin{tikzpicture}[spy using outlines = {rectangle, magnification=2.0, connect spies}]
    \begin{semilogyaxis}[ 
            footnotesize, width=\columnwidth, height=.67\columnwidth,    
            xmin=0, xmax=6, xtick={0,1,...,6},
            xlabel=$\frac{E_b}{N_0}$ (dB), ylabel=FER,   
            grid=both, grid style={gray!30},
            tick align=outside, tickpos=left, 
            legend pos=south west, 
            legend cell align={left},
            /pgfplots/table/ignore chars={|},
            mark options={solid},
        ]

       \addplot[mark=square , Paired-7, semithick]  table[x=Eb/N0, y=FER] {data/BCH/63_45/FER_ORBGRAND_FER_LW63_HW4.txt};
       \addplot[mark=square , Paired-6, semithick]  table[x=Eb/N0, y=FER] {data/BCH/63_45/FER_ORBGRAND_FER_LW63_HW5.txt};
       \addplot[mark=square, Paired-1, semithick]  table[x=Eb/N0, y=FER] {data/BCH/63_45/FER_ORBGRAND_FER_LW63_HW6.txt};
       \addplot[mark=square  , Paired-3, semithick]  table[x=Eb/N0, y=FER] {data/BCH/63_45/FER_ORBGRAND_FER_LW63_HW8.txt};
       \addplot[mark=square  , Paired-5, semithick]  table[x=Eb/N0, y=FER] {data/BCH/63_45/FER_ORBGRAND_FER_LW63_HW63.txt};
       \addplot[mark=square  , Paired-11, semithick]  table[x=Eb/N0, y=FER] {data/BCH/63_45/FER_ORBGRAND_FER_LW2016_HW63.txt};

       \coordinate (spypoint1) at (axis cs:5.5,1.3e-5);
      \coordinate (magnifyglass1) at (axis cs:4.0,1.2e-5);

        \legend{{} {\tiny{($LW_\text{max}$ = 63, $HW_\text{max}$ = 4)}},
        {} {\tiny{($LW_\text{max}$ = 63, $HW_\text{max}$ = 5})},
        {} {\tiny{($LW_\text{max}$ = 63, $HW_\text{max}$ = 6})},
        {} {\tiny{($LW_\text{max}$ = 63, $HW_\text{max}$ = 8})},
        {} {\tiny{($LW_\text{max}$ = 63, $HW_\text{max}$ = 63)}},
        {} {\tiny{($LW_\text{max}$ = 2016, $HW_\text{max}$ = 63)}},
         } 
    \end{semilogyaxis}
\end{tikzpicture}  
\vspace*{-2em}
\caption{\label{fig:fer_bch_63_45}Comparison of decoding performance of ORBGRAND ($LW_\text{max}$,$HW_\text{max}$) decoding of BCH Code (63,45).}
\vspace*{-1em}
\end{figure}  



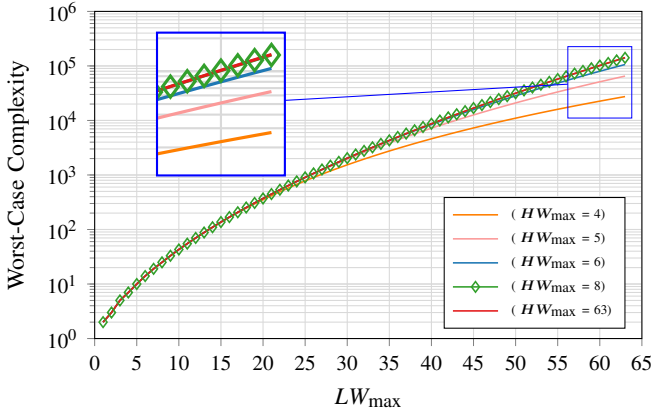
\begin{figure}[!t]
 \centering
\begin{tikzpicture}[spy using outlines = {rectangle, magnification=2.0, connect spies}]
    \begin{semilogyaxis}[ 
            footnotesize, width=\columnwidth, height=.67\columnwidth, 
            xmin=0, xmax=65, xtick={0,5,...,65}, 
            ymin=1, ymax = 1e6,  
            xlabel=$LW_\text{max}$, ylabel=Worst-Case Complexity,   
            grid=both, grid style={gray!30},
            tick align=outside, tickpos=left, 
            legend pos=south east, 
            legend cell align={left},
            /pgfplots/table/ignore chars={|},
            mark options={solid},
        ]

       \addplot[ Paired-7, semithick]  table[x=Param, y=Comp] {data/BCH/63_45/FER_ORBGrand_WC_Latency_LW63_HW4.txt};
       \addplot[ Paired-6, semithick]  table[x=Param, y=Comp] {data/BCH/63_45/FER_ORBGrand_WC_Latency_LW63_HW5.txt};
       \addplot[ Paired-1, semithick]  table[x=Param, y=Comp] {data/BCH/63_45/FER_ORBGrand_WC_Latency_LW63_HW6.txt};
       \addplot[mark=diamond, Paired-3, semithick]  table[x=Param, y=Comp] {data/BCH/63_45/FER_ORBGrand_WC_Latency_LW63_HW8.txt};
       \addplot[ Paired-5, semithick]  table[x=Param, y=Comp] {data/BCH/63_45/FER_ORBGrand_WC_Latency_LW63_HW63.txt};

       \coordinate (spypoint1) at (axis cs:60,0.5e5);
      \coordinate (magnifyglass1) at (axis cs:15,0.2e5);

        \legend{{} {\tiny{( $HW_\text{max}$ = 4)}},
        {} {\tiny{( $HW_\text{max}$ = 5})},
        {} {\tiny{( $HW_\text{max}$ = 6})},
        {} {\tiny{( $HW_\text{max}$ = 8})},
        {} {\tiny{( $HW_\text{max}$ = 63)}}
         }
         \spy [blue, width=1.7cm, height=1.9cm] on (spypoint1) in node[fill=white] at (magnifyglass1); 
    \end{semilogyaxis}
\end{tikzpicture}  
\vspace*{-2em}
\caption{\label{fig:comp_bch_63_45}Maximum number of queries (worst-case complexity) comparison for ORBGRAND decoding of BCH Code (63,45).}
\vspace*{-1em}
\end{figure}  


\section{Enhancing the error decoding performance of ORBGRAND}
In this section, we will look at techniques for boosting ORBGRAND's decoding performance to match the ML decoding performance of SGRAND. We begin by analysing the effect of ORBGRAND's parameters on the decoding performance, and then we propose a list-based technique to improve ORBGRAND's decoding performance. The proposed List-GRAND algorithm introduces parameters that can be tweaked to match the ML decoding performance of SGRAND as well as the target decoding performance and complexity budget of a specific application.

\subsection{Parametric analysis of ORBGRAND}\label{sec:ORBGRANDcomp}

$LW_\text{max}$ and $HW_\text{max}$ are two important ORBGRAND parameters that impact both decoding performance and the maximum number of codebook membership queries required (the worst-case complexity) by ORBGRAND. The impact of parameters ($LW_\text{max}$, $HW_\text{max}$) on the decoding performance and the worst-case complexity of ORBGRAND for decoding BCH code (63,45) with BPSK modulation over an AWGN channel is depicted in Figure \ref{fig:fer_bch_63_45} and Fig. \ref{fig:comp_bch_63_45} respectively. The performance of ORBGRAND decoding is improved by increasing the values of the parameters $LW_\text{max}$ and $HW_\text{max}$; however, as shown in Fig. \ref{fig:fer_bch_63_45}, the worst-case complexity also increases.

\begin{algorithm}[t]
\caption{\label{alg:Lgrand}LGRAND Algorithm}
    \DontPrintSemicolon
    \SetAlgoVlined  
    \SetKwData{e}{$\bm{e}$}
    \SetKwData{p}{$\bm{p}$}
    \SetKwData{s}{$\bm{S}$}
    \SetKwData{ind}{$\bm{ind}$}
    \SetKwData{LLR}{$\bm{y}$}
    \SetKwData{sortSet}{$[\bm{r},\bm{ind}]$}
    \SetKwData{estm}{$\hat{\bm{u}}$}
    \SetKwData{ginv}{$\bm{G}^{-1}$}
    \SetKwData{LW}{${LW_\text{max}}$}
    \SetKwData{HW}{${HW_\text{max}}$}
    \SetKwData{yhat}{$\hat{\bm{y}}$}
    \SetKwData{chat}{$\hat{\bm{c}}$}
    \SetKwData{chatfinal}{$\hat{\bm{c}}_{final}$}
    \KwIn{\LLR, $\bm{H}$, \ginv, \LW, \HW, $\delta$}
    \KwOut{\estm}
    \SetKwFunction{RecursiveComputeLLRs}{recursiveComputeLLRs}
    \SetKwFunction{DecodeRONE}{decodeR1}
    \SetKwFunction{HammingWeight}{HammingWeight}
    \SetKwFunction{RDecodeRONE}{redecodeR1}
    \SetKwFunction{DecodeRZERO}{decodeR0}
    \SetKwFunction{Find}{findCandidate}
    \SetKwFunction{new}{genErrorPatternMaxHammingWt}
    \SetKwFunction{intPartition}{generateAllIntPartitions}
    \SetKwFunction{Sort}{sortChannelObservationValues}
    \SetKwFunction{List}{addToList}
    \eIf{$\bm{H} \cdot\yhat^\top == \bm{0}$}
{\KwRet{$\estm \leftarrow\yhat\cdot\ginv$}}
{
    $\ind \leftarrow$ \Sort{\LLR} \tcp*[r]{$\lvert{\bm{y}}_i\rvert\leq\lvert{\bm{y}}_j\rvert~~\forall i < j$}
    $\e \leftarrow \bm{0}$; 
    $\Lambda \leftarrow \LW$; 
    $\Delta \leftarrow \HW$; \;
     $\mathcal{L} \leftarrow \emptyset$; \; 
    \For{$i \gets 1$ to $\Lambda$}{
        $\s \leftarrow$ \intPartition{i}\tcp*[r]{$(\lambda_1, \lambda_2, \ldots, \lambda_P) \vdash i$}
        \ForAll{$\bm{l}$ in \s}{
          $\e \leftarrow$ \new{$\bm{l}$,\ind, $\Delta$}\tcp*[r]{$\HammingWeight{\e} \leq \Delta$}
          \If{$\bm{H} \cdot(\yhat \oplus \e)^\top == \bm{0}$} {
            $\chat \leftarrow \yhat \oplus \e$\;
            \List{$\mathcal{L}$,\chat}\;
            \If{$\Lambda == \LW$} {
            $\Lambda \leftarrow min(i+\delta$,~$LW_\text{max}$) \;
            $\Delta \leftarrow \HammingWeight{\e}$\;
            } 
            }
        }
        }
        $\chatfinal \leftarrow \argmax\limits_{\chat\in \mathcal{L}} p(\bm{y}|\chat)$\;
        $\estm \leftarrow \chatfinal\cdot\ginv$\;
        \KwRet{\estm}
    }
\end{algorithm}

\begin{figure}
  \centering
  \includegraphics[width=1\linewidth]{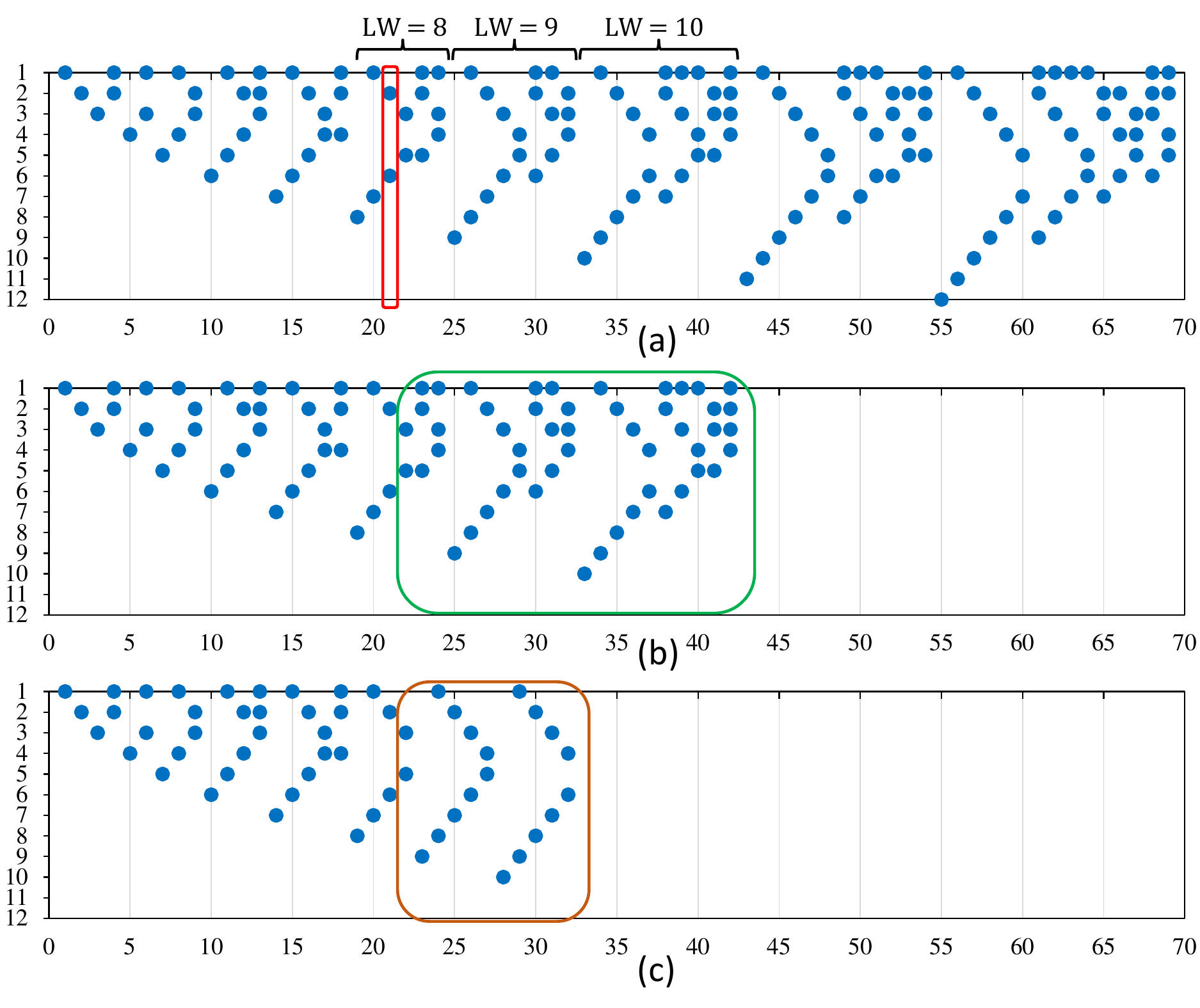}
  \vspace*{-2em}
  \caption{Test Error Pattern (TEP) generation for LGRAND for ($n=12$, $LW_{max}$=12, $HW_{max}$=4) (a) Upper: Codebook membership criterion (\ref{eq:constraint}) satisfied by $21^{st}$ TEP ($\bm{e}$) with $LW$=8 and $HW=2$ (red rectangle). (b) Middle: Checking additional TEPs for LGRAND. ($\delta=2$) (green rectangle). (c) Bottom: Restricting HW of additional TEPs to $\leq2$ ($\delta=2$ and $\Delta=Hamming~Weight(\bm{e})$) (brown rectangle).}.
  \label{fig:TEP_sch_lgrand} 
  \vspace*{-1em}
\end{figure}

\subsection{Proposed List-GRAND (LGRAND)} \label{sec:List-GRAND}
Algorithm \ref{alg:Lgrand} describes the proposed LGRAND decoding approach. The inputs of LGRAND are identical to those of ORBGRAND, with the exception of an extra parameter $\delta$ (threshold for logistic weight). Unlike ORBGRAND, which terminates decoding as soon as any vector ($\hat{\bm{y}}~\oplus~\bm{e}$) fulfils the codebook membership criterion (\ref{eq:constraint}), LGRAND generates a list ($\mathcal{L}$) of estimated codewords ($\hat{\bm{c}}$) and selects the most likely one ($\argmax\limits_{\hat{\bm{c}}\in \mathcal{L}} p(\bm{y}|\hat{\bm{c}})$) as the final estimated codeword $\hat{\bm{c}}_{final}$.

LGRAND proceeds similarly to ORBGRAND by sorting $\bm{y}$ in ascending order of absolute value ($\lvert{\bm{y}}_i\rvert\leq\lvert{\bm{y}}_j\rvert~~\forall i < j$), and the corresponding indices are recorded in a permutation vector denoted by $\bm{ind}$ (line 4). Following this, integer partitions of a logistic weight $i$ $(\forall i \in [0, \Lambda], \text{where } \Lambda =LW_\text{max})$ are generated. Then, LGRAND generates TEPs ($\bm{e}$) by using the generated integer partitions and the TEPs are ordered using the permutation vector $\bm{ind}$ (line 10). Please note that the Hamming weight of the generated TEPs is restricted to $\leq\Delta$ ($\Delta$ is initialized to $HW_{max}$ (line 5)). These TEPs are then applied to $\hat{\bm{y}}$ to check for codebook membership criterion (\ref{eq:constraint}). Whenever a vector $\hat{\bm{y}}~\oplus~\bm{e}$  meets the codebook membership criterion (\ref{eq:constraint}), LGRAND adds the vector $\hat{\bm{y}}~\oplus~\bm{e}$ to the list $\mathcal{L}$ (line 13).

Fig. \ref{fig:TEP_sch_lgrand}(a) depicts the ORBGRAND TEPs for parameters $n=12$, $LW_{max}=12$, $HW_{max}=4$ and $\delta=2$. Suppose that the $21^{st}$ TEP, which corresponds to an integer partition of $8$ ($LW=8$) and has a Hamming weight of $2$, fulfils the codebook membership criterion (\ref{eq:constraint}). Rather than stopping the decoding process, LGRAND checks additional TEPs corresponding to $LW = 9$ and $LW = 10$ ($\delta=2$), as illustrated in Fig. \ref{fig:TEP_sch_lgrand} (b). If any of these additional TEPs meet codebook membership constraint (\ref{eq:constraint}), they are added to the List $\mathcal{L}$, and the most likely codeword is chosen as the final codeword  $\hat{\bm{c}}_{final}$ (line 18).

To reduce the number of generated additional TEPs, the maximum Hamming weight of the additional TEPs is restricted to the Hamming weight of the first TEP ($\Delta=Hamming~Weight(\bm{e})$) that satisfied the codebook membership criterion (\ref{eq:constraint}) when combined with $\hat{\bm{y}}$ (line 16). Limiting the Hamming weight of additional TEPs implies that only TEPs $(\bm{e})$ with Hamming weights $\leq\Delta$ will be generated, as shown in Fig. \ref{fig:TEP_sch_lgrand}(c).

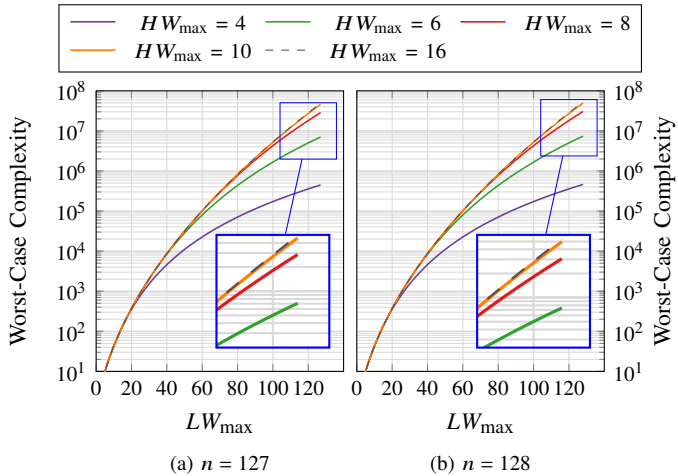
\begin{figure}[!t]
  \centering
  \begin{tikzpicture}[spy using outlines = {rectangle, magnification=2.0, connect spies}]
    \begin{groupplot}[group style={group name=fer_queries, group size= 2 by 1, horizontal sep=5pt, vertical sep=5pt},
      footnotesize,
      height=.6\columnwidth,  width=0.55\columnwidth,
      xlabel=$LW_\text{max}$ ,
      xmin=0, xmax=140, xtick={0,20,...,120}, 
      ymode=log,
      tick align=inside,
      grid=both, grid style={gray!30},
      /pgfplots/table/ignore chars={|},
      ]
      \nextgroupplot[ylabel= Worst-Case Complexity, ytick pos=left, y label style={at={(axis description cs:-0.225,.5)},anchor=south},ymin=1e1, ymax = 1e8]

      \addplot[ Paired-9, semithick]  table[x=Param, y=Comp] {data/BCH/WC_Lat/FER_ListGrand_WC_Latency_LW127_HW4.txt};\label{gp:plot1_WC_127}
      \addplot[ Paired-3, semithick]  table[x=Param, y=Comp] {data/BCH/WC_Lat/FER_ListGrand_WC_Latency_LW127_HW6.txt};\label{gp:plot2_WC_127}
      \addplot[ Paired-5, semithick]  table[x=Param, y=Comp] {data/BCH/WC_Lat/FER_ListGrand_WC_Latency_LW127_HW8.txt};\label{gp:plot3_WC_127}
      \addplot[ Paired-7, semithick]  table[x=Param, y=Comp] {data/BCH/WC_Lat/FER_ListGrand_WC_Latency_LW127_HW10.txt};\label{gp:plot4_WC_127}
      \addplot[mark=none, Paired-13, dashed]  table[x=Param, y=Comp] {data/BCH/WC_Lat/FER_ListGrand_WC_Latency_LW127_HW16.txt};\label{gp:plot7_WC_127}

      \coordinate (spypoint1) at (axis cs:120,1e7);
      \coordinate (magnifyglass1) at (axis cs:100,1e3);

      \coordinate (top) at (rel axis cs:0,1);
\nextgroupplot[ylabel=Worst-Case Complexity, ytick pos=right,y label style={at={(axis description cs:1.325,.5)},anchor=south},ymin=1e1, ymax = 1e8]

\addplot[ Paired-9, semithick]  table[x=Param, y=Comp] {data/CRC/WC_LAT/FER_ListGrand_WC_Latency_LW128_HW4.txt};
      \addplot[ Paired-3, semithick]  table[x=Param, y=Comp] {data/CRC/WC_LAT/FER_ListGrand_WC_Latency_LW128_HW6.txt};
      \addplot[ Paired-5, semithick]  table[x=Param, y=Comp] {data/CRC/WC_LAT/FER_ListGrand_WC_Latency_LW128_HW8.txt};
      \addplot[ Paired-7, semithick]  table[x=Param, y=Comp] {data/CRC/WC_LAT/FER_ListGrand_WC_Latency_LW128_HW10.txt};
      \addplot[mark=none, Paired-13, dashed]  table[x=Param, y=Comp] {data/CRC/WC_LAT/FER_ListGrand_WC_Latency_LW128_HW16.txt};

      \coordinate (spypoint2) at (axis cs:120,1.2e7);
      \coordinate (magnifyglass2) at (axis cs:100,1e3);
      
      \coordinate (bot) at (rel axis cs:1,0);
      \spy [blue, width=1.5cm, height=1.5cm] on (spypoint1) in node[fill=white] at (magnifyglass1); 
      \spy [blue, width=1.5cm, height=1.5cm] on (spypoint2) in node[fill=white] at (magnifyglass2); 
    \end{groupplot}
    \node[below = 1cm of fer_queries c1r1.south] {\footnotesize (a) $n=127$ };
    \node[below = 1cm of fer_queries c2r1.south] {\footnotesize (b) $n=128$ };
    \path (top|-current bounding box.north) -- coordinate(legendpos) (bot|-current bounding box.north);
    \matrix[
    matrix of nodes,
    anchor=south,
    draw,
    inner sep=0.2em,
    draw
    ]at(legendpos)
    { \ref{gp:plot1_WC_127}& \footnotesize{ $HW_\text{max}$ = 4} &[1pt]
    \ref{gp:plot2_WC_127}& \footnotesize{ $HW_\text{max}$ = 6} &[1pt]
    \ref{gp:plot3_WC_127}& \footnotesize{ $HW_\text{max}$ = 8} \\
    \ref{gp:plot4_WC_127}& \footnotesize{ $HW_\text{max}$ = 10} &[1pt]
    \ref{gp:plot7_WC_127}& \footnotesize{ $HW_\text{max}$ = 16} \\
       }; 

  \end{tikzpicture}
  \vspace*{-2em}
  \caption{\label{fig:comp_128_127}Worst-case complexity for ORBGRAND and LGRAND decoding of linear block codes of length $n$.}
 \vspace*{-1em}
\end{figure}

\begin{figure}[!t]
  \centering
  \begin{tikzpicture}[spy using outlines = {rectangle, magnification=2.0, connect spies}]
    \begin{groupplot}[group style={group name=fer_queries, group size= 2 by 1, horizontal sep=5pt, vertical sep=5pt},
      footnotesize,
      height=.6\columnwidth,  width=0.55\columnwidth,
      xlabel=$\frac{E_b}{N_0}$ (dB),
      xmin=0, xmax=9, xtick={0,1,...,8},
      ymode=log,
      tick align=inside,
      grid=both, grid style={gray!30},
      /pgfplots/table/ignore chars={|},
      ]

      \nextgroupplot[ylabel= FER, ytick pos=left, y label style={at={(axis description cs:-0.225,.5)},anchor=south},ymin=3e-8, ymax = 2]
      \addplot[mark=diamond* , Paired-7, semithick]  table[x=Eb/N0, y=FER] {data/BCH/127_113/BCH_N127_113_BM.txt};\label{gp:plot1_bch_127_113}
      \addplot[mark=triangle, Paired-1, semithick]  table[x=Eb/N0, y=FER] {data/BCH/127_113/BCH_N127_113_ORBGRAND.txt};\label{gp:plot3_bch_127_113}
      \addplot[mark=square  , Paired-3, semithick]  table[x=Eb/N0, y=FER] {data/BCH/127_113/BCH_N127_113_SGRAND.txt}  ;\label{gp:plot4_bch_127_113}
      \addplot[mark=o       , Paired-6, semithick]  table[x=Eb/N0, y=FER] {data/BCH/127_113/BCH_N127_113_ORBGRAND_LW_96_HW_8.txt}    ;\label{gp:plot5_bch_127_113}
      \addplot[mark=pentagon, Paired-5, semithick]  table[x=Eb/N0, y=FER] {data/BCH/127_113/BCH_N127_113_ML.txt}      ;\label{gp:plot6_bch_127_113};\
      \addplot[mark=diamond, Paired-12, semithick]  table[x=Eb/N0, y=FER] {data/BCH/127_113/BCH_N127_113_LISTGRAND_25_LW96_HW8.txt};\label{gp:plot11_bch_127_113}
      \addplot[mark=+       , Paired-9, semithick]  table[x=Eb/N0, y=FER] {data/BCH/127_113/FER_N127_113_ILW_10_4.txt};\label{gp:plot9_bch_127_113}
      \addplot[mark=star       , Paired-11, semithick]  table[x=Eb/N0, y=FER] {data/BCH/127_113/FER_N127_113_ILW_10_3.txt};\label{gp:plot8_bch_127_113}
      \coordinate (top) at (rel axis cs:0,1);

      \coordinate (spypoint1) at (axis cs:7.8,0.1e-6);
      \coordinate (magnifyglass1) at (axis cs:2.6,1.1e-5);

      \coordinate (spypoint2) at (axis cs:15.4,0.25e-6);
      \coordinate (magnifyglass2) at (axis cs:10.3,0.15e-4);

      \nextgroupplot[ylabel=Avg. Queries, ytick pos=right,y label style={at={(axis description cs:1.325,.5)},anchor=south},ymin=1, ymax = 5e5]
      \addplot[mark=diamond* , Paired-7, semithick]  table[x=Eb/N0, y=Tests/f] {data/BCH/127_113/BCH_N127_113_BM.txt};
      \addplot[mark=triangle, Paired-1, semithick]  table[x=Eb/N0, y=Tests/f] {data/BCH/127_113/BCH_N127_113_ORBGRAND.txt};
      \addplot[mark=square  , Paired-3, semithick]  table[x=Eb/N0, y=Tests/f] {data/BCH/127_113/BCH_N127_113_SGRAND.txt}  ;
      \addplot[mark=o       , Paired-6, semithick]  table[x=Eb/N0, y=Tests/f] {data/BCH/127_113/BCH_N127_113_ORBGRAND_LW_96_HW_8.txt}   ;
      
      \addplot[mark=diamond       , Paired-12, semithick]  table[x=Eb/N0, y=Tests/f] {data/BCH/127_113/BCH_N127_113_LISTGRAND_25_LW96_HW8.txt}    ;
      \addplot[mark=+       , Paired-9, semithick]  table[x=Eb/N0, y=Comp] {data/BCH/127_113/FER_N127_113_ILW_10_4.txt};
      \addplot[mark=star       , Paired-11, semithick]  table[x=Eb/N0, y=Comp] {data/BCH/127_113/FER_N127_113_ILW_10_3.txt};
      \coordinate (bot) at (rel axis cs:1,0);
    \end{groupplot}
    \node[below = 1cm of fer_queries c1r1.south] {\footnotesize (a) : FER};
    \node[below = 1cm of fer_queries c2r1.south] {\footnotesize (b) : Avg. Queries};
    \path (top|-current bounding box.north) -- coordinate(legendpos) (bot|-current bounding box.north);
    \matrix[
    matrix of nodes,
    anchor=south,
    draw,
    inner sep=0.2em,
    draw
    ]at(legendpos)
    {
      \ref{gp:plot6_bch_127_113}& \footnotesize ML  &[1pt]
      
      \ref{gp:plot5_bch_127_113}& \footnotesize ORBGRAND ($LW_\text{max}$=96,$HW_\text{max}= 8$) \\
      \ref{gp:plot3_bch_127_113}& \footnotesize ORBGRAND &[1pt] 
      \ref{gp:plot11_bch_127_113}& \footnotesize LGRAND ($LW_\text{max}$=96,$HW_\text{max}= 8$,$\delta = 25$) \\
      
      \ref{gp:plot4_bch_127_113}& \footnotesize SGRAND &[1pt]
      \ref{gp:plot8_bch_127_113}& \footnotesize ORBGRAND-ILWO (Queries$_\text{max}$=$10^{3}$) \\
      \ref{gp:plot1_bch_127_113}& \footnotesize GRANDAB &[1pt]
      \ref{gp:plot9_bch_127_113}& \footnotesize ORBGRAND-ILWO (Queries$_\text{max}$=$10^{4}$) \\
      }; 
      \spy [blue, width=1.5cm, height=0.9cm] on (spypoint1) in node[fill=white] at (magnifyglass1);
  \end{tikzpicture}
  \vspace*{-2em}
  \caption{\label{fig:grand_bch_2}Comparison of decoding performance and average complexity of different GRAND variants for BCH(127, 113) code.}
   \vspace*{-1em}
\end{figure}
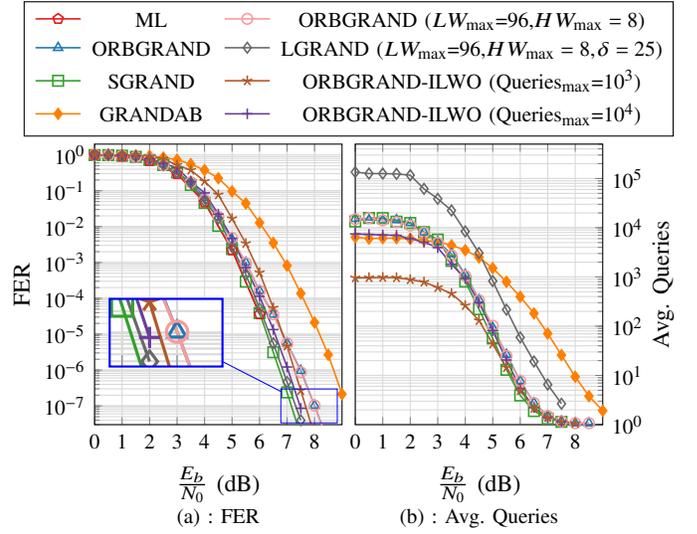


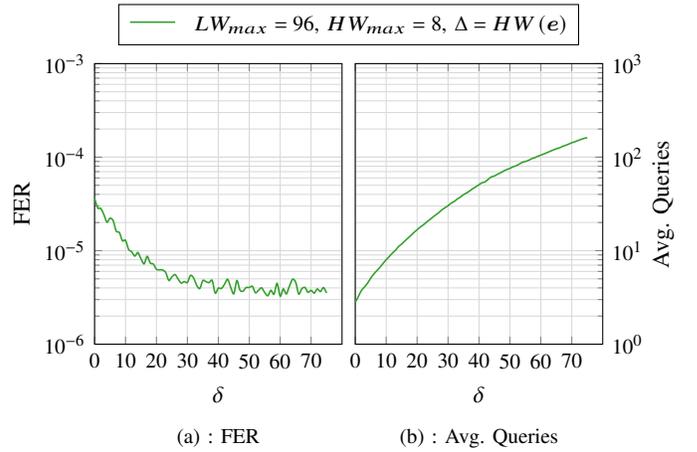
\begin{figure}[!t]
  \centering
  \begin{tikzpicture}[spy using outlines = {rectangle, magnification=2.0, connect spies}]
    \begin{groupplot}[group style={group name=fer_queries, group size= 2 by 1, horizontal sep=5pt, vertical sep=5pt},
      footnotesize,
      height=.6\columnwidth,  width=0.55\columnwidth,
      xlabel=$\delta$ ,
      xmin=0, xmax=80, xtick={0,10,...,70}, 
      ymode=log,
      tick align=inside,
      grid=both, grid style={gray!30},
      /pgfplots/table/ignore chars={|},
      ]

      \nextgroupplot[ylabel= FER, ytick pos=left, y label style={at={(axis description cs:-0.225,.5)},anchor=south},ymin=1e-6, ymax = 1e-3]
      \addplot[smooth,Paired-3, semithick]  table[x=Param, y=FER] {data/BCH/127_113/FER_ListGrand_findParameter_127_113_LW96_HW8_ConfineLW.txt};\label{gp:plot2_param_127_113}

      \coordinate (top) at (rel axis cs:0,1);

      \nextgroupplot[ylabel=Avg. Queries, ytick pos=right,y label style={at={(axis description cs:1.325,.5)},anchor=south},ymin=1, ymax = 1e3]
      \addplot[smooth,Paired-3, semithick]  table[x=Param, y=Comp] {data/BCH/127_113/FER_ListGrand_findParameter_127_113_LW96_HW8_ConfineLW.txt};

      \coordinate (bot) at (rel axis cs:1,0);
    \end{groupplot}
    \node[below = 1cm of fer_queries c1r1.south] {\footnotesize (a) : FER };
    \node[below = 1cm of fer_queries c2r1.south] {\footnotesize (b) : Avg. Queries };
    \path (top|-current bounding box.north) -- coordinate(legendpos) (bot|-current bounding box.north);
    \matrix[
    matrix of nodes,
    anchor=south,
    draw,
    inner sep=0.2em,
    draw
    ]at(legendpos)
    {
      \ref{gp:plot2_param_127_113}& \footnotesize{ $LW_{max}=96$, $HW_{max}=8$, $\Delta=HW(\bm{e})$} \\
       }; 
  \end{tikzpicture}
  \vspace*{-2em}
  \caption{\label{fig:param_bch_127_113}Parametric analysis of LGRAND ($LW_{max}$, $HW_{max}$, $\delta$) for BCH (127,113) code (at $\frac{E_b}{N_0}=6.5$ dB).}
 \vspace*{-1em}
\end{figure}

\subsection{Parametric analysis of LGRAND }
The LGRAND technique introduces the parameter $\delta$, that influences both the decoding performance and the average complexity (average number of codebook membership queries (TEPs)) of the algorithm. It should be noted that the worst-case complexity of LGRAND is the same as that of ORBGRAND, because the worst-case complexity of ORBGRAND and LGRAND is dependent on the parameters $LW_\text{max}$ and $HW_\text{max}$, which are same for both algorithms (section \ref{sec:ORBGRANDcomp}). The worst-case complexity for ORBGRAND and LGRAND decoding of linear block codes with lengths $n=127$ and $n=128$ is shown in Fig. \ref{fig:comp_128_127}.

Figure \ref{fig:grand_bch_2} compares the decoding performance and average complexity of different GRAND variants for decoding BCH code $(127,113)$. Furthermore, the ML decoding performance results \cite{kaiserslautern} are included for reference. The parameter $AB=2$ is chosen for the GRANDAB hard-input decoder. As shown in Fig. \ref{fig:grand_bch_2} (a), SGRAND (Queries$_\text{max}$=$10^{6}$) outperforms ORBGRAND in decoding performance by $0.8$ dB at the target FER of $10^{-7}$. However, with the appropriate choice of parameter $\delta$, the proposed LGRAND technique can bridge the decoding performance gap between SGRAND and ORBGRAND as shown in Fig. \ref{fig:grand_bch_2} (a).

The effect of changing the value of $\delta$ on both the decoding performance and the average computational complexity for LGRAND decoding of BCH code $(127,113)$ at $\frac{E_b}{N_0}=6.5$ dB is depicted in Fig. \ref{fig:param_bch_127_113}. As shown in Fig. \ref{fig:param_bch_127_113} (a), increasing the value of parameter $\delta$ improves decoding performance (FER at $\frac{E_b}{N_0}=6.5$ dB), but it also increases average computational complexity, as shown in Fig. \ref{fig:param_bch_127_113} (b). As a consequence, the appropriate value of parameter $\delta$  can be chosen to strike a balance between decoding performance and average complexity. At a target FER of $10^{-7}$, LGRAND with parameters ($LW_{max}=96$, $HW_{max}=8$, $\delta=25$) achieves error decoding performance comparable to SGRAND and outperforms ORBGRAND by $0.75$dB, as shown in Fig \ref{fig:grand_bch_2} (a).

\subsubsection*{Analyzing average list size ($|\mathcal{L}|_{avg}$)}
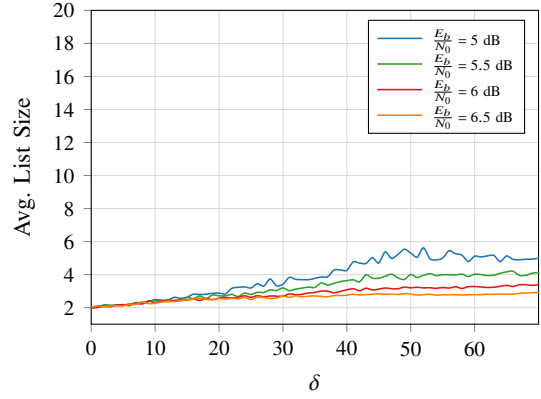
\begin{figure}[!t]
\begin{tikzpicture}
    \begin{axis}[
            footnotesize, width=0.85\columnwidth, height=.65\columnwidth,    
            xmin=0, xmax=70, xtick={0,10,...,60}, 
            ymin=1,  ymax=20,
            xlabel=$\delta$, ylabel=Avg. List Size,  
            grid=both, grid style={gray!30},
            tick align=outside, tickpos=left, 
            legend pos=north east, legend cell align={left},
            legend style={nodes={scale=0.7, transform shape}},
            /pgfplots/table/ignore chars={|},
        ]

      \addplot[smooth,Paired-1, semithick]  table[x=param, y=ListSize] {data/BCH/127_113/FER_ListGrand_96_8_AvgListSize_5dB.txt};\label{gp:plotListSize_1}
      \addplot[smooth,Paired-3, semithick]  table[x=param, y=ListSize] {data/BCH/127_113/FER_ListGrand_96_8_AvgListSize_5_5dB.txt};\label{gp:plotListSize_2}
      \addplot[smooth,Paired-5, semithick]  table[x=param, y=ListSize] {data/BCH/127_113/FER_ListGrand_96_8_AvgListSize_6dB.txt};\label{gp:plotListSize_3}
      \addplot[smooth,Paired-7, semithick]  table[x=param, y=ListSize] {data/BCH/127_113/FER_ListGrand_96_8_AvgListSize_6_5dB.txt};\label{gp:plotListSize_4}
         \legend{{} {$\frac{E_b}{N_0} = 5$ dB},{} {$\frac{E_b}{N_0} = 5.5$ dB},{} {$\frac{E_b}{N_0} = 6$ dB},{} {$\frac{E_b}{N_0} = 6.5$ dB},
         }
    \end{axis}
\end{tikzpicture}  
\caption{\label{fig:Avg_List_Size} Average list size ($|\mathcal{L}|_{avg}$) for LGRAND ($LW_{max}=96$, $HW_{max}=8$, $\Delta=HW(\bm{e})$) decoding for BCH code $(127,113)$.}
\end{figure}

The effect of parameter $\delta$ on the average list size ($|\mathcal{L}|_{avg}$) for LGRAND ($LW_{max}=96$, $HW_{max}=8$, $\delta=25$) decoding for BCH code $(127,113)$ at different $\frac{E_b}{N_0}$ values is depicted in Fig. \ref{fig:Avg_List_Size}. Please note that at least 100 errors are collected for each value of parameter $\delta$, and the average List size is plotted in Fig. \ref{fig:Avg_List_Size} for various $\frac{E_b}{N_0}$ value. As seen in Fig. \ref{fig:Avg_List_Size}, the average list size increases as the value of $\delta$ increases for all $\frac{E_b}{N_0}$ values. However, as shown in Fig. \ref{fig:param_bch_127_113}, increasing the value of $\delta$ improves decoding performance at the expense of average computational complexity.

\subsubsection*{Suboptimality Analysis}
The ORBGRAND applies the TEPs in a predetermined logistic weight order as discussed in section \ref{sec:TEPGen}, and it is obvious from the performance difference between the ORBGRAND and SGRAND decoder—which applies the TEPs in an optimal (ML) order—that this logistic weight order is not the optimal schedule. By generating a list of potential candidates during the decoding process and choosing the most likely candidate, the proposed LGRAND improves the decoding performance of ORBGRAND.

\begin{figure}[!t]
\begin{tikzpicture}
\begin{axis} [ybar, bar width=7pt,
width=0.85\columnwidth, height=.65\columnwidth, 
enlargelimits=0.15,  
    xlabel=\footnotesize{$\frac{E_b}{N_0}$ (dB)}, 
    ylabel=\footnotesize{Suboptimality count},  
    ymode=log,
    tick align=inside,
    grid=both, grid style={gray!30},
    xmin=0.5, xmax=7, xtick={0,1,...,7},
    ymin=0, ymax=10^3,
]
\addplot table[x=Eb/N0, y=Sub_Opt]{data/BCH/127_113/FER_ListGrand_LW96_HW8_25_SubOptimalRatio.txt};
\end{axis}
 
\end{tikzpicture}
  \vspace*{-1em}
  \caption{\label{fig:SubOptimal} Suboptimality count for LGRAND ($LW_\text{max}$=96,$HW_\text{max}= 8$,$\delta = 25$) decoding of BCH code $(127,113)$.}
 \vspace*{-1em}
\end{figure}
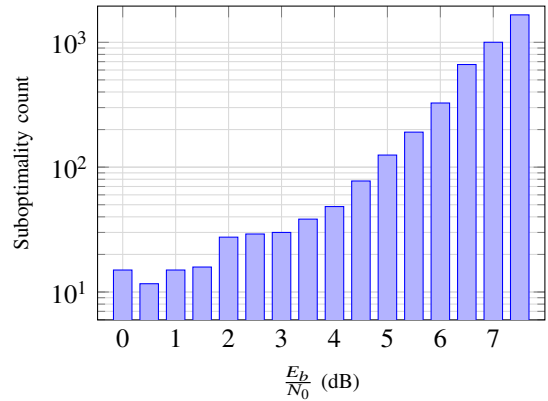

\textit{Suboptimality count} refers to the number of instances where the most likely candidate is not the first on the list ($\mathcal{L}$). Please note that if the selected candidate is the first on the list ($\mathcal{L}$), the LGRAND will perform similar to the ORBGRAND.
Fig. \ref{fig:SubOptimal} illustrates the suboptimality count for LGRAND ($LW_{max}=96$, $HW_{max}=8$, $\delta=25$) decoding of BCH Code $(127,113)$. It should be noted that at least 100 errors were captured at each $\frac{E_b}{N_0}$ point. As observed in Fig. \ref{fig:SubOptimal}, the suboptimality count increases with higher $\frac{E_b}{N_0}$ values, revealing the ORBGRAND's suboptimality and providing an explanation for the performance improvement achieved by the proposed LGRAND.

\subsection{Comparison with enhanced ORBGRAND TEP scheduling schemes}

Recently, an improved ORBGRAND TEP scheduling technique was presented in \cite{carloILW}; this approach generates TEPs using Improved Logistics Weight Order (ILWO) as opposed to the conventional logistic weight order \cite{duffy2020ordered}. The TEPs with lower Hamming weights are given precedence in the improved logistic weight order, whereas the TEPs with higher Hamming weights are penalized. The parameter Queries$_\text{max}$, which indicates the maximum number of queries allowed, also influences the computational complexity of the ILWO-ORBGRAND as well as the decoding performance \cite{carloILW}. We refer the reader to \cite{carloILW} for more details on the proposed ILWO TEP schedule and performance/complexity tradeoffs. Figure \ref{fig:grand_bch_2} (a) illustrates the ORBGRAND decoder with the ILWO TEP schedule \cite{carloILW}, which outperforms the traditional ORBGRAND decoder \cite{duffy2020ordered} by $0.3-0.55$ dB at the target FER of $10^{-7}$. To strike a balance between the decoding performance and computational complexity, appropriate values of Queries$_\text{max}$ can be selected.

Please note that baseline ORBGRAND and LGRAND TEP schedules allow parallel online computation of TEPs using simple $n\times(n-k)$-bit shift registers and a network of XOR gates, as illustrated in Section \ref{sec:LGRAND-VLSI}. Thus, the proposed LGRAND presents itself as a viable option for a hardware-friendly solution to achieve ML decoding performance. In a similar way, the ORBGRAND-ILWO \cite{carloILW} can also be implemented in hardware leveraging a network of XOR gates and shift-registers with a few minor modifications.


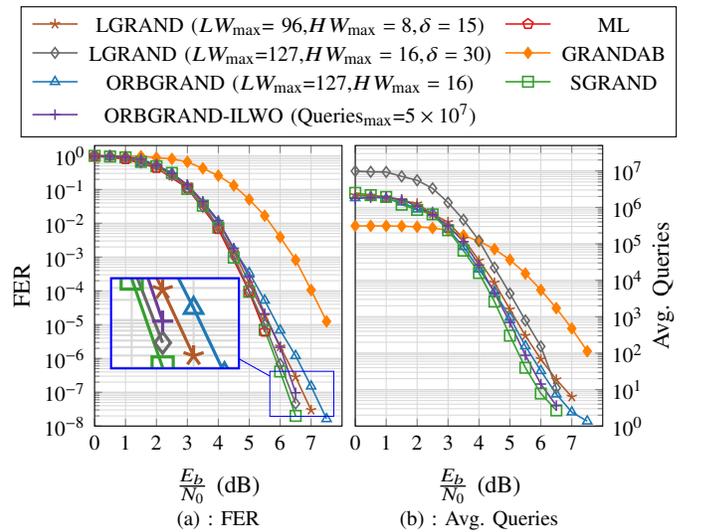
\begin{figure}[!t]
  \centering
  \begin{tikzpicture}[spy using outlines = {rectangle, magnification=2.0, connect spies}]
    \begin{groupplot}[group style={group name=fer_queries, group size= 2 by 1, horizontal sep=5pt, vertical sep=5pt},
      footnotesize,
      height=.6\columnwidth,  width=0.55\columnwidth,
      xlabel=$\frac{E_b}{N_0}$ (dB),
      xmin=0, xmax=8, xtick={0,1,...,7},
      ymode=log,
      tick align=inside,
      grid=both, grid style={gray!30},
      /pgfplots/table/ignore chars={|},
      ]

      \nextgroupplot[ylabel= FER, ytick pos=left, y label style={at={(axis description cs:-0.225,.5)},anchor=south},ymin=1e-8, ymax = 2]
      \addplot[mark=diamond* , Paired-7, semithick]  table[x=Eb/N0, y=FER] {data/BCH/127_106/BCH_N127_106_BM.txt};\label{gp:plot1_bch_127_106}
      \addplot[mark=triangle, Paired-1, semithick]  table[x=Eb/N0, y=FER] {data/BCH/127_106/BCH_N127_106_ORBGRAND_LW_128_HW_16.txt};\label{gp:plot3_bch_127_106}
      \addplot[mark=square  , Paired-3, semithick]  table[x=Eb/N0, y=FER] {data/BCH/127_106/BCH_N127_106_SGRAND_5_10_7.txt}  ;\label{gp:plot4_bch_127_106}
      \addplot[mark=pentagon, Paired-5, semithick]  table[x=Eb/N0, y=FER] {data/BCH/127_106/BCH_N127_106_ML.txt}      ;\label{gp:plot6_bch_127_106}

      \addplot[mark=star, Paired-11, semithick]  table[x=Eb/N0, y=FER] {data/BCH/127_106/BCH_N127_106_LISTGRAND_15.txt};\label{gp:plot9_bch_127_106}

      \addplot[mark=diamond, Paired-12, semithick]  table[x=Eb/N0, y=FER] {data/BCH/127_106/BCH_N127_106_LISTGRAND_30.txt};\label{gp:plot11_bch_127_106}

      \addplot[mark=+       , Paired-9, semithick]  table[x=Eb/N0, y=FER] {data/BCH/127_106/FER_N127_106_ILW_5_10_7.txt};\label{gp:plot8_bch_127_106}

      \coordinate (top) at (rel axis cs:0,1);

      \coordinate (spypoint1) at (axis cs:6.7,0.9e-7);
      \coordinate (magnifyglass1) at (axis cs:2.6,1.1e-5);

      \nextgroupplot[ylabel=Avg. Queries, ytick pos=right,y label style={at={(axis description cs:1.325,.5)},anchor=south},ymin=1, ymax = 5e7]
      \addplot[mark=diamond* , Paired-7, semithick]  table[x=Eb/N0, y=Tests/f] {data/BCH/127_106/BCH_N127_106_BM.txt};
      \addplot[mark=triangle, Paired-1, semithick]  table[x=Eb/N0, y=Tests/f] {data/BCH/127_106/BCH_N127_106_ORBGRAND_LW_128_HW_16.txt};
      \addplot[mark=square  , Paired-3, semithick]  table[x=Eb/N0, y=Tests/f] {data/BCH/127_106/BCH_N127_106_SGRAND_5_10_7.txt}  ;
      \addplot[mark=star       , Paired-11, semithick]  table[x=Eb/N0, y=Tests/f] {data/BCH/127_106/BCH_N127_106_LISTGRAND_15.txt}    ;

      \addplot[mark=diamond       , Paired-12, semithick]  table[x=Eb/N0, y=Tests/f] {data/BCH/127_106/BCH_N127_106_LISTGRAND_30.txt}    ;

      \addplot[mark=+       , Paired-9, semithick]  table[x=Eb/N0, y=Tests/f] {data/BCH/127_106/FER_N127_106_ILW_5_10_7.txt};
      \coordinate (bot) at (rel axis cs:1,0);

    \end{groupplot}
    \node[below = 1cm of fer_queries c1r1.south] {\footnotesize (a) : FER};
    \node[below = 1cm of fer_queries c2r1.south] {\footnotesize (b) : Avg. Queries};
    \path (top|-current bounding box.north) -- coordinate(legendpos) (bot|-current bounding box.north);
    \matrix[
    matrix of nodes,
    anchor=south,
    draw,
    inner sep=0.2em,
    draw
    ]at(legendpos)
    {
      \ref{gp:plot9_bch_127_106}& \footnotesize LGRAND ($LW_\text{max}$= 96,$HW_\text{max}=8$,$\delta = 15$) &[1pt] 
      \ref{gp:plot6_bch_127_106}& \footnotesize ML  \\
      
      \ref{gp:plot11_bch_127_106}& \footnotesize LGRAND ($LW_\text{max}$=127,$HW_\text{max}= 16$,$\delta = 30$)&[1pt]
      \ref{gp:plot1_bch_127_106}& \footnotesize GRANDAB\\
      
      \ref{gp:plot3_bch_127_106}& \footnotesize ORBGRAND ($LW_\text{max}$=127,$HW_\text{max}= 16$)  &[1pt]
      \ref{gp:plot4_bch_127_106}& \footnotesize SGRAND \\
      \ref{gp:plot8_bch_127_106}& \footnotesize ORBGRAND-ILWO (Queries$_\text{max}$=$5\times10^{7}$) \\
      }; 
      \spy [blue, width=1.7cm, height=1.2cm] on (spypoint1) in node[fill=white] at (magnifyglass1);
  \end{tikzpicture}
  \vspace*{-2em}
  \caption{\label{fig:grand_bch_1} Comparison of decoding performance and average complexity of different GRAND variants for BCH $(127, 106)$ code.}
 \vspace*{-1em}
\end{figure}
\begin{figure}[!t]
  \centering
  \begin{tikzpicture}[spy using outlines = {rectangle, magnification=2.0, connect spies}]
    \begin{groupplot}[group style={group name=fer_queries, group size= 2 by 1, horizontal sep=5pt, vertical sep=5pt},
      footnotesize,
      height=.6\columnwidth,  width=0.55\columnwidth,
      xlabel=$\frac{E_b}{N_0}$ (dB),
      xmin=0, xmax=8, xtick={0,1,...,7},
      ymode=log,
      tick align=inside,
      grid=both, grid style={gray!30},
      /pgfplots/table/ignore chars={|},
      ]

      \nextgroupplot[ylabel= FER, ytick pos=left, y label style={at={(axis description cs:-0.225,.5)},anchor=south},ymin=3e-8, ymax = 2]
      \addplot[mark=diamond* , Paired-7, semithick]  table[x=Eb/N0, y=FER] {data/CRC/128_104/CRC_N128_104_GRANDAB.txt};\label{gp:plot1_crc_128_104}
      \addplot[mark=triangle, Paired-1, semithick]  table[x=Eb/N0, y=FER] {data/CRC/128_104/CRC_N128_104_ORBGRAND_LW_128_HW_16.txt};\label{gp:plot3_crc_128_104}
      \addplot[mark=square  , Paired-3, semithick]  table[x=Eb/N0, y=FER] {data/CRC/128_104/CRC_N128_104_SGRAND_5_10_7.txt}  ;\label{gp:plot4_crc_128_104}
      \addplot[mark=diamond, Paired-12, semithick]  table[x=Eb/N0, y=FER] {data/CRC/128_104/CRC_N128_104_LISTGRAND_30.txt};\label{gp:plot11_crc_128_104}
      \addplot[mark=+       , Paired-9, semithick]  table[x=Eb/N0, y=FER] {data/CRC/128_104/CRC_N128_104_ILW_5_10_7.txt}; \label{gp:plot8_crc_128_104}

      \coordinate (top) at (rel axis cs:0,1);

      \coordinate (spypoint1) at (axis cs:6.3,0.2e-6);
      \coordinate (magnifyglass1) at (axis cs:2.6,1.1e-5);

      \nextgroupplot[ylabel=Avg. Queries, ytick pos=right,y label style={at={(axis description cs:1.325,.5)},anchor=south},ymin=1, ymax = 5e7]
      \addplot[mark=diamond* , Paired-7, semithick]  table[x=Eb/N0, y=Tests/f] {data/CRC/128_104/CRC_N128_104_GRANDAB.txt};
      \addplot[mark=triangle, Paired-1, semithick]  table[x=Eb/N0, y=Tests/f] {data/CRC/128_104/CRC_N128_104_ORBGRAND_LW_128_HW_16.txt};
      \addplot[mark=square  , Paired-3, semithick]  table[x=Eb/N0, y=Tests/f] {data/CRC/128_104/CRC_N128_104_SGRAND_5_10_7.txt}  ;
      \addplot[mark=diamond       , Paired-12, semithick]  table[x=Eb/N0, y=Tests/f] {data/CRC/128_104/CRC_N128_104_LISTGRAND_30.txt}    ;
      \addplot[mark=+       , Paired-9, semithick]  table[x=Eb/N0, y=Tests/f] {data/CRC/128_104/CRC_N128_104_ILW_5_10_7.txt}; 

      \coordinate (bot) at (rel axis cs:1,0);
    \end{groupplot}
    \node[below = 1cm of fer_queries c1r1.south] {\footnotesize (a) : FER};
    \node[below = 1cm of fer_queries c2r1.south] {\footnotesize (b) : Avg. Queries};
    \path (top|-current bounding box.north) -- coordinate(legendpos) (bot|-current bounding box.north);
    \matrix[
    matrix of nodes,
    anchor=south,
    draw,
    inner sep=0.2em,
    draw
    ]at(legendpos)
    {
      \ref{gp:plot3_crc_128_104}& \footnotesize ORBGRAND ($LW_\text{max}=128$,$HW_\text{max}=16$) &[1pt] 
      \ref{gp:plot1_crc_128_104}& \footnotesize GRANDAB \\
      \ref{gp:plot11_crc_128_104}& \footnotesize LGRAND ($LW_\text{max}=128$,$HW_\text{max}=16$,$\delta = 30$) &[1pt] 
      \ref{gp:plot4_crc_128_104}& \footnotesize SGRAND \\
      \ref{gp:plot8_crc_128_104}& \footnotesize ORBGRAND-ILWO (Queries$_\text{max}$=$5\times10^{7}$) \\
       }; 
      \spy [blue, width=1.4cm, height=1.0cm] on (spypoint1) in node[fill=white] at (magnifyglass1);
  \end{tikzpicture}
  \vspace*{-2em}
  \caption{\label{fig:grand_crc_128_104}Comparison of decoding performance and average complexity of different GRAND variants for CRC Code (128,104).}
   \vspace*{-1em}
\end{figure}
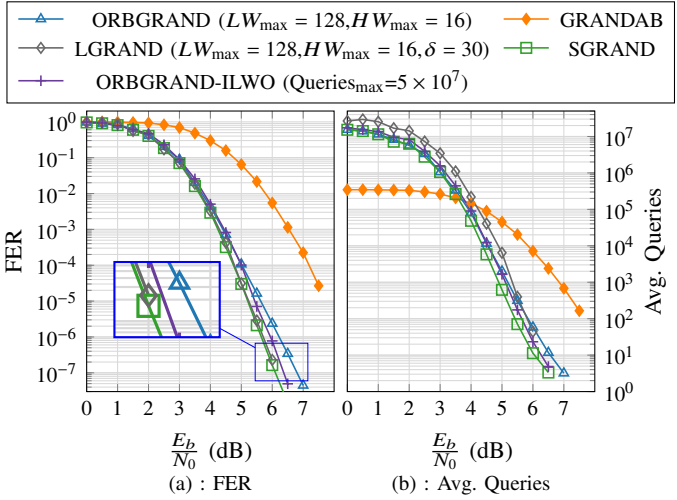

\begin{figure}[!t]
  \centering
  \begin{tikzpicture}[spy using outlines = {rectangle, magnification=2.0, connect spies}]
    \begin{groupplot}[group style={group name=fer_queries, group size= 2 by 1, horizontal sep=5pt, vertical sep=5pt},
      footnotesize,
      height=.6\columnwidth,  width=0.55\columnwidth,
      xlabel=$\frac{E_b}{N_0}$ (dB),
      xmin=0, xmax=9, xtick={0,1,...,9},
      ymode=log,
      tick align=inside,
      grid=both, grid style={gray!30},
      /pgfplots/table/ignore chars={|},
      ]

      \nextgroupplot[ylabel= FER, ytick pos=left, y label style={at={(axis description cs:-0.225,.5)},anchor=south},ymin=3e-8, ymax = 2]
      \addplot[mark=diamond* , Paired-7, semithick]  table[x=Eb/N0, y=FER] {data/CRC/128_112/CRC_N128_112_GRANDAB.txt};\label{gp:plot1_crc_128_112}
      \addplot[mark=triangle, Paired-1, semithick]  table[x=Eb/N0, y=FER] {data/CRC/128_112/CRC_N128_112_ORBGRAND_LW_96_HW_8.txt};\label{gp:plot3_crc_128_112}
      \addplot[mark=square  , Paired-3, semithick]  table[x=Eb/N0, y=FER] {data/CRC/128_112/CRC_N128_112_SGRAND_3_10_6.txt}  ;\label{gp:plot4_crc_128_112}
      \addplot[mark=diamond, Paired-12, semithick]  table[x=Eb/N0, y=FER] {data/CRC/128_112/CRC_N128_112_LISTGRAND_24.txt};\label{gp:plot11_crc_128_112}
      \addplot[mark=+       , Paired-9, semithick]  table[x=Eb/N0, y=FER] {data/CRC/128_112/CRC_N128_112_ILW_3_10_6.txt}; \label{gp:plot8_crc_128_112}

      \coordinate (top) at (rel axis cs:0,1);

      \coordinate (spypoint1) at (axis cs:7.8,0.1e-6);
      \coordinate (magnifyglass1) at (axis cs:2.6,1.1e-5);

      \nextgroupplot[ylabel=Avg. Queries, ytick pos=right,y label style={at={(axis description cs:1.325,.5)},anchor=south},ymin=1, ymax = 1e6]
      \addplot[mark=diamond* , Paired-7, semithick]  table[x=Eb/N0, y=Tests/f] {data/CRC/128_112/CRC_N128_112_GRANDAB.txt};
      \addplot[mark=triangle, Paired-1, semithick]  table[x=Eb/N0, y=Tests/f] {data/CRC/128_112/CRC_N128_112_ORBGRAND_LW_96_HW_8.txt};
      \addplot[mark=square  , Paired-3, semithick]  table[x=Eb/N0, y=Tests/f] {data/CRC/128_112/CRC_N128_112_SGRAND_3_10_6.txt}  ;
      \addplot[mark=diamond       , Paired-12, semithick]  table[x=Eb/N0, y=Tests/f] {data/CRC/128_112/CRC_N128_112_LISTGRAND_24.txt}    ;
      \addplot[mark=+       , Paired-9, semithick]  table[x=Eb/N0, y=Tests/f] {data/CRC/128_112/CRC_N128_112_ILW_3_10_6.txt}; 

      \coordinate (bot) at (rel axis cs:1,0);
    \end{groupplot}
    \node[below = 1cm of fer_queries c1r1.south] {\footnotesize (a) : FER};
    \node[below = 1cm of fer_queries c2r1.south] {\footnotesize (b) : Avg. Queries};
    \path (top|-current bounding box.north) -- coordinate(legendpos) (bot|-current bounding box.north);
    \matrix[
    matrix of nodes,
    anchor=south,
    draw,
    inner sep=0.2em,
    draw
    ]at(legendpos)
    {
      \ref{gp:plot3_crc_128_112}& \footnotesize ORBGRAND ($LW_\text{max}=96$,$HW_\text{max}=8$) &[1pt] 
      \ref{gp:plot1_crc_128_112}& \footnotesize GRANDAB \\      
      \ref{gp:plot11_crc_128_112}& \footnotesize LGRAND ($LW_\text{max}=96$,$HW_\text{max}=8$,$\delta = 24$) &[1pt] 
      \ref{gp:plot4_crc_128_112}& \footnotesize SGRAND \\
      \ref{gp:plot8_crc_128_112}& \footnotesize ORBGRAND-ILWO (Queries$_\text{max}$=$3\times10^{6}$) \\
       }; 
      \spy [blue, width=1.0cm, height=0.8cm] on (spypoint1) in node[fill=white] at (magnifyglass1);
  \end{tikzpicture}
  \vspace*{-2em}
  \caption{\label{fig:grand_crc_128_112}Comparison of decoding performance and average complexity of different GRAND variants for CRC Code (128,112).}
   \vspace*{-1em}
\end{figure}
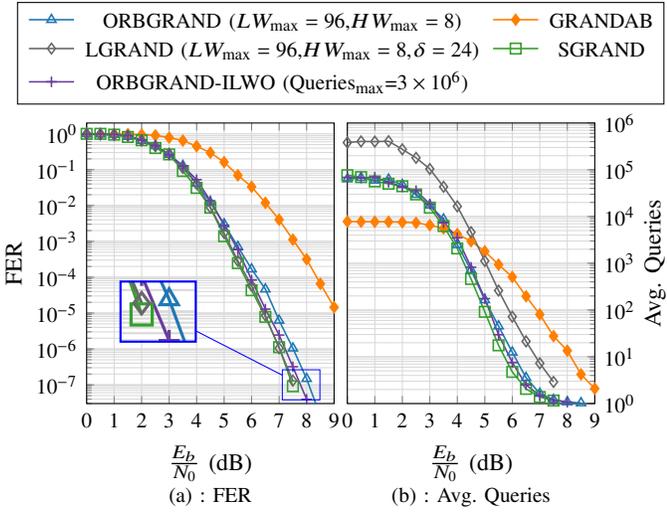
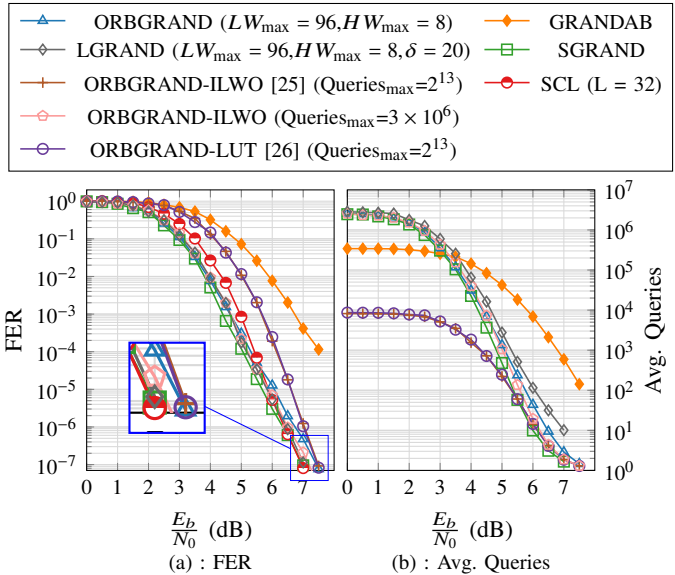
\begin{figure}[!t]
  \centering
  \begin{tikzpicture}[spy using outlines = {rectangle, magnification=2.0, connect spies}]
    \begin{groupplot}[group style={group name=fer_queries, group size= 2 by 1, horizontal sep=5pt, vertical sep=5pt},
      footnotesize,
      height=.6\columnwidth,  width=0.55\columnwidth,
      xlabel=$\frac{E_b}{N_0}$ (dB),
      xmin=0, xmax=8, xtick={0,1,...,7},
      ymode=log,
      tick align=inside,
      grid=both, grid style={gray!30},
      /pgfplots/table/ignore chars={|},
      ]

      \nextgroupplot[ylabel= FER, ytick pos=left, y label style={at={(axis description cs:-0.225,.5)},anchor=south},ymin=7e-8, ymax = 2]
      \addplot[mark=diamond* , Paired-7, semithick]  table[x=Eb/N0, y=FER] {data/Polar/128_105/POLAR_N128_105_GRANDAB.txt};\label{gp:plot1_p}
      \addplot[mark=triangle, Paired-1, semithick]  table[x=Eb/N0, y=FER] {data/Polar/128_105/POLAR_N128_105_ORBGRAND_LW_96_HW_8.txt};\label{gp:plot3_p}
      \addplot[mark=square  , Paired-3, semithick]  table[x=Eb/N0, y=FER] {data/Polar/128_105/POLAR_N128_105_SGRAND_3_10_6.txt}  ;\label{gp:plot4_p}
      \addplot[mark=halfcircle*  , Paired-5, semithick]  table[x=Eb/N0, y=FER] {data/Polar/128_105/POLAR_N128_105_L_32.txt}      ;\label{gp:plot6_p};
      \addplot[mark=+  , Paired-11 , semithick]  table[x=Eb/N0, y=FER] {data/Polar/128_105/FER_N128_105_ILW_8192.txt}; \label{gp:plot7_p}
      \addplot[mark=o  , Paired-9 , semithick]  table[x=Eb/N0, y=FER] {data/Polar/128_105/FER_N128_105_LUT_ILW_8192.txt}; \label{gp:plot8_p}
      \addplot[mark=pentagon  , Paired-6 , semithick]  table[x=Eb/N0, y=FER] {data/Polar/128_105/FER_N128_105_ILW_3_10_6.txt}; \label{gp:plot5_p}
      \addplot[mark=diamond, Paired-12, semithick]  table[x=Eb/N0, y=FER] {data/Polar/128_105/POLAR_N128_105_LISTGRAND_20_96_8.txt};\label{gp:plot11_p}

      \coordinate (top) at (rel axis cs:0,1);

      \coordinate (spypoint1) at (axis cs:7.2,0.15e-6);
      \coordinate (magnifyglass1) at (axis cs:2.6,1.1e-5);

      \nextgroupplot[ylabel=Avg. Queries, ytick pos=right,y label style={at={(axis description cs:1.325,.5)},anchor=south},ymin=1, ymax = 1e7]
      \addplot[mark=diamond* , Paired-7, semithick]  table[x=Eb/N0, y=Tests/f] {data/Polar/128_105/POLAR_N128_105_GRANDAB.txt};
      \addplot[mark=triangle, Paired-1, semithick]  table[x=Eb/N0, y=Tests/f] {data/Polar/128_105/POLAR_N128_105_ORBGRAND_LW_96_HW_8.txt};
      \addplot[mark=square  , Paired-3, semithick]  table[x=Eb/N0, y=Tests/f] {data/Polar/128_105/POLAR_N128_105_SGRAND_3_10_6.txt}  ;
      \addplot[mark=+  , Paired-11 , semithick]  table[x=Eb/N0, y=Tests/f] {data/Polar/128_105/FER_N128_105_ILW_8192.txt}; 
      \addplot[mark=o  , Paired-9 , semithick]  table[x=Eb/N0, y=Tests/f] {data/Polar/128_105/FER_N128_105_LUT_ILW_8192.txt}; 
      \addplot[mark=pentagon  , Paired-6 , semithick]  table[x=Eb/N0, y=Tests/f] {data/Polar/128_105/FER_N128_105_ILW_3_10_6.txt};
      \addplot[mark=diamond       , Paired-12, semithick]  table[x=Eb/N0, y=Tests/f] {data/Polar/128_105/POLAR_N128_105_LISTGRAND_20_96_8.txt}    ;

      \coordinate (bot) at (rel axis cs:1,0);
    \end{groupplot}
    \node[below = 1cm of fer_queries c1r1.south] {\footnotesize (a) : FER};
    \node[below = 1cm of fer_queries c2r1.south] {\footnotesize (b) : Avg. Queries};
    \path (top|-current bounding box.north) -- coordinate(legendpos) (bot|-current bounding box.north);
    \matrix[
    matrix of nodes,
    anchor=south,
    draw,
    inner sep=0.2em,
    draw
    ]at(legendpos)
    {
      \ref{gp:plot3_p}& \footnotesize ORBGRAND ($LW_\text{max}=96$,$HW_\text{max}=8$) &[1pt]  
      \ref{gp:plot1_p}& \footnotesize GRANDAB \\
      \ref{gp:plot11_p}& \footnotesize LGRAND ($LW_\text{max}=96$,$HW_\text{max}=8$,$\delta = 20$) &[1pt] 
      \ref{gp:plot4_p}& \footnotesize SGRAND \\
      \ref{gp:plot7_p}& \footnotesize ORBGRAND-ILWO \cite{carloILW} (Queries$_\text{max}$=$2^{13}$) &[1pt]
      \ref{gp:plot6_p}& \footnotesize SCL (L = 32) \\ 
      \ref{gp:plot5_p}& \footnotesize ORBGRAND-ILWO (Queries$_\text{max}$=$3\times10^{6}$) \\
      \ref{gp:plot8_p}& \footnotesize ORBGRAND-LUT \cite{carloORB} (Queries$_\text{max}$=$2^{13}$) \\
       }; 
      \spy [blue, width=1.0cm, height=1.2cm] on (spypoint1) in node[fill=white] at (magnifyglass1);
  \end{tikzpicture}
  \vspace*{-2em}
  \caption{\label{fig:grand_polar}Comparison of decoding performance and average complexity of different GRAND variants for Polar Code (128,105+11).}
   \vspace*{-1em}
\end{figure}

\section{Performance Evaluation}\label{sec:performance} 

In this section, we evaluate the proposed LGRAND in terms of decoding performance and computational complexity for distinct classes of channel codes (BCH, CA-Polar, and CRC). Fig. \ref{fig:grand_bch_1} (a) compares the FER performance of LGRAND with different variants of GRAND for decoding BCH code $(127,106)$.  In addition, the ML decoding \cite{kaiserslautern} results are included for reference. Please note that the maximum number of queries (worst-case complexity) for the LGRAND and ORBGRAND decoders is $4.93\times10^7$ with codes of length 127 , which correspond to the parameters $LW_\text{max}$=127 and $HW_\text{max}= 16$ (Fig. \ref{fig:comp_128_127}). Furthermore, for the numerical simulation results shown in Fig. \ref{fig:grand_bch_1}, the parameter Queries$_\text{max}$=$5\times10^{7}$ is used for SGRAND decoder and $AB = 3$ for GRANDAB decoder.

As demonstrated in Fig. \ref{fig:grand_bch_1} (a), while both soft-input variants of GRAND (ORBGRAND and SGRAND) outperform the hard-input GRANDAB, SGRAND achieves the ML performance. The proposed LGRAND (with different parameter settings) outperforms ORBGRAND in decoding performance by $0.25-0.7$dB at a target FER of $10^{-7}$. Furthermore, as explained in the preceding subsection, these parameters can be tweaked to match SGRAND's ML decoding performance. For the BCH code $(127,106)$, LGRAND with parameters $LW_{max}=127$, $HW_{max}=16$, and $\delta=30$ results in a decoding performance gain of $0.7$dB over ORBGRAND at a target FER of $10^{-7}$ as depicted in Fig. \ref{fig:grand_bch_1} (a). Additionally, as demonstrated in Fig. \ref{fig:grand_bch_1} (a), the improved TEP schedule ORBGRAND-ILWO (Queries$_\text{max}$=$5\times10^{7}$) \cite{carloILW} outperforms the baseline ORBGRAND by $~0.6$ dB at a target FER of $10^{-7}$. However, as stated in the preceding section, the proposed LGRAND, which is based on the ORBGRAND TEP schedule, is a suitable choice for parallel hardware implementation since it supports online parallel TEP generation using shift registers and a network of XOR gates.

The average computational complexity for different GRAND variants is shown in Fig. \ref{fig:grand_bch_1} (b). Despite the fact that SGRAND requires the fewest queries of any GRAND variant, as explained in the previous section, it is not suitable for parallel hardware implementation. As a result, comparing the number of queries required by the proposed LGRAND to the number of queries required by ORBGRAND is reasonable because both are equally suitable for parallel hardware implementation.

Fig. \ref{fig:grand_crc_128_104} and Fig. \ref{fig:grand_crc_128_112} compare LGRAND decoding performance, as well as average computational complexity, with other GRAND variants for Cyclic Redundancy Check (CRC) codes \cite{Peterson61}. CRC codes are typically used to detect errors in communication systems and to assist list-based channel code decoders in selecting the final candidate codeword. On the other hand, CRC codes can also be used for error correction using the GRAND algorithm. The concept of using CRC codes for error correction with GRAND decoding was presented in \cite{GRANDAB-VLSI} and expanded on in \cite{CRCGrand}.  For CRC code $(128,104)$ and CRC code $(128,112)$, the generator polynomial are \texttt{0xB2B117} and \texttt{0x1021} respectively. 

The worst-case complexity of both LGRAND and ORBGRAND decoders, corresponding to parameters ($LW_\text{max}$=128, $HW_\text{max}= 16$) and ($LW_\text{max}$=96 and $HW_\text{max}= 8$), with codes of length 128 is $5.33\times10^7$ and $3.10\times10^6$ (Fig. \ref{fig:comp_128_127}) respectively. Furthermore, Queries$_\text{max}$=$5\times10^{7}$ is employed for SGRAND and $AB=3$ is selected for GRANDAB in the numerical simulation results displayed in Fig. \ref{fig:grand_crc_128_104}. Similarly, for the simulation results shown in Fig. \ref{fig:grand_crc_128_112}, Queries$_\text{max}$=$3\times10^{6}$ and $AB=2$ for the SGRAND and GRANDAB decoders, respectively.

At the target FER of $10^{-7}$, LGRAND ($LW_{max}=128$, $HW_{max}=16$, $\delta=30$) achieves similar decoding performance to SGRAND (Queries$_\text{max}$=$5\times10^{7}$) for the CRC code $(128,104)$ shown in Fig. \ref{fig:grand_crc_128_104}. Similarly, with the CRC code $(128,112)$  shown in Fig. \ref{fig:grand_crc_128_112}, LGRAND ($LW_{max}=96$, $HW_{max}=8$, $\delta=24$) achieves SGRAND (Queries$_\text{max}$=$3\times10^{6}$) decoding performance ($0.5$dB gain over ORBGRAND at the target FER of $10^{-7}$). Furthermore, as illustrated in  Fig. \ref{fig:grand_crc_128_104} and Fig. \ref{fig:grand_crc_128_112}, the proposed LGRAND is compared with ORBGRAND-ILWO. At the target FER of $10^{-7}$, as shown in  Fig. \ref{fig:grand_crc_128_104} and Fig. \ref{fig:grand_crc_128_112}, the proposed LGRAND performs better than the ORBGRAND-ILWO \cite{carloILW} by $0.1-0.2$ dB.

Fig. \ref{fig:grand_polar} compares the proposed LGRAND's decoding performance, as well as the required average number of queries, with different variants of GRAND for decoding 5G NR CA-polar code (128,105+11). Furthermore, the decoding performance of state-of-the-art soft-input decoder such as the CA-SCL decoder \cite{Tal15,LLR-List} is included for reference. Please note that for LGRAND decoding of CA-polar code (128,105+11), the CRC bits are not used to select the most likely candidate from the list ($\mathcal{L}$). Instead, we select the most likely candidate ($\argmax\limits_{\hat{\bm{c}}\in \mathcal{L}} p(\bm{y}|\hat{\bm{c}})$) from the list using the maximum likelihood criterion (Section \ref{sec:List-GRAND}). 

The worst-case complexity of the ORBGRAND and LGRAND decoder, which corresponds to parameters $LW_\text{max}$=96 and $HW_\text{max}= 8$, is $3.10\times10^6$ (Fig. \ref{fig:comp_128_127}) for the numerical simulation results depicted in Fig. \ref{fig:grand_polar}. Furthermore, SGRAND employs Queries$_\text{max}$=$3\times10^{6}$ and the GRANDAB decoder employs $AB = 3$.

The LGRAND decoder is also compared with enhanced ORGBGRAND TEP schedules, ORBGRAND-ILWO \cite{carloILW} and Look-Up-Table (LUT) assisted Fixed Latency ORBGRAND decoder \cite{carloORB} (F.L ORBGRAND), using the same $(128,105)$ polar code shown in Fig. \ref{fig:grand_polar}. At a target FER of $10^{-7}$, LGRAND ($LW_{max}=96$, $HW_{max}=8$, $\delta=20$) achieves a decoding performance similar to SGRAND and outperforms traditional ORBGRAND \cite{duffy2020ordered} as well as enhanced TEP schedule ORBGRAND (Queries$_\text{max}$=$2^{13}$) \cite{carloILW,carloORB} by $\sim0.5$dB as shown in Fig. \ref{fig:grand_polar}. 

To conclude, LGRAND's parameters ($LW_{max}$, $HW_{max}$, $\delta$) can be appropriately chosen for channel codes of different classes (BCH, CA-Polar, and CRC) to achieve ML decoding performance. Furthermore, the complexity overhead for different LGRAND parameter choices can be explored further in order to strike a balance between decoding performance requirements and the complexity/latency budget for a target application. 

\begin{figure}
  \centering
  \includegraphics[width=0.95\linewidth]{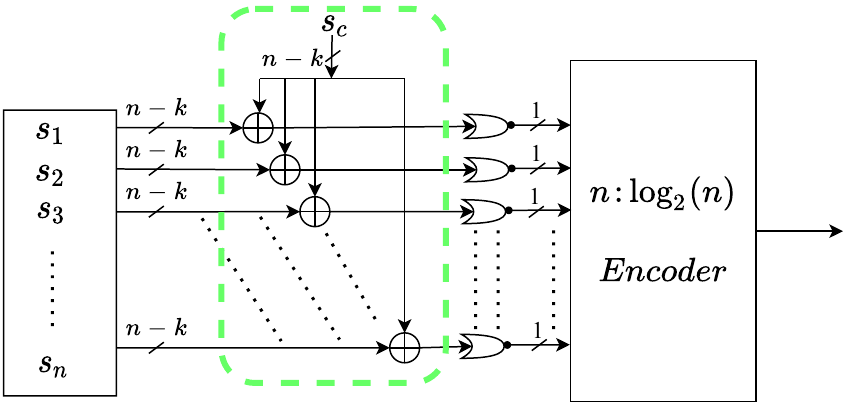}
  \caption{VLSI architecture for checking error patterns with Hamming weight of $1$ ($\bm{s}_i = \bm{H}\cdot\mathds{1}_i^\top$, $i \in \llbracket 1\isep n \rrbracket$).}
  \label{fig:arch_1bit} 
  \vspace*{-1em}
\end{figure}

\begin{figure*}
\centering
  \includegraphics[width=0.85\linewidth]{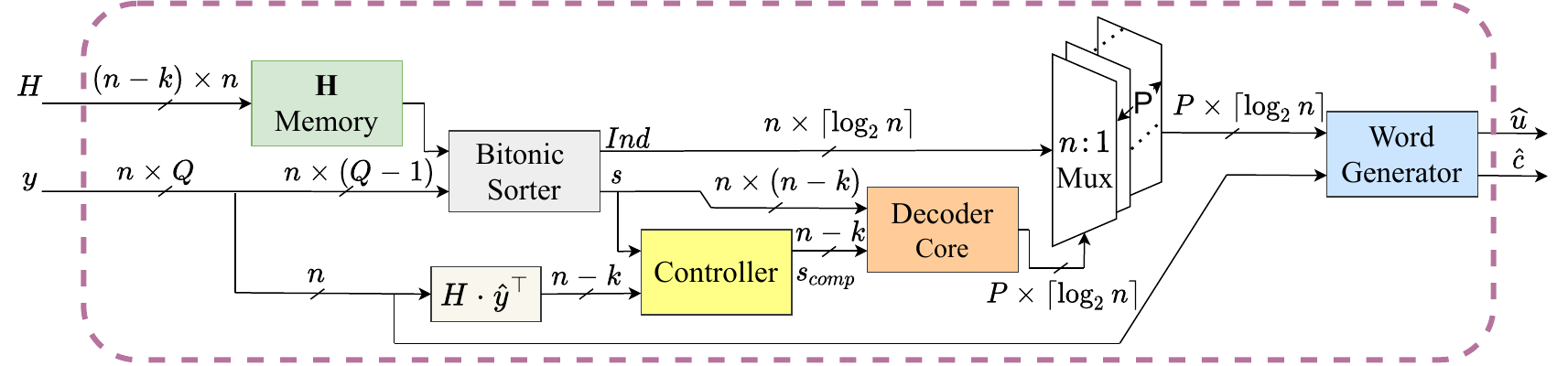}
  \caption{VLSI Architecture for ORBGRAND \cite{ORBGRAND-TVLSI}.}
  \label{fig:ORBarch}
  \vspace*{-1em}
\end{figure*}

\section{VLSI Architecture for LGRAND}

This section describes the proposed VLSI architecture for LGRAND. For $(n,k)$ linear block codes, VLSI architectures for GRANDAB and ORBGRAND were proposed in \cite{GRANDAB-VLSI} and \cite{ORBGRAND-TVLSI}. Without going into details, we will briefly explain the techniques used in \cite{GRANDAB-VLSI} and \cite{ORBGRAND-TVLSI} to generate TEPs with Hamming weights $\geq1$. By using the parity check matrix ($\bm{H}$) and the received vector $\hat{\bm{y}}$ from the channel, the test error pattern with Hamming weight of $1$ ($\bm{e}=\mathds{1}_i$, $i \in \llbracket 1\isep n \rrbracket$) can be checked for codebook membership as 
\begin{equation}
\bm{H} \cdot(\hat{\bm{y}} \oplus \mathds{1}_i)^\top=\bm{H}\cdot\hat{\bm{y}}^\top \oplus \bm{H}\cdot\mathds{1}_i^\top, 
\label{eq:one-bit-constraint}
\end{equation}
where $\bm{H}\cdot\hat{\bm{y}}^\top$ (denoted as $\bm{s}_c$) is the $(n-k)$-bits syndrome associated with the received
vector $\hat{\bm{y}}$, and $\bm{H}\cdot\mathds{1}_i^\top$ (denoted as $\bm{s}_i$) is the
$(n-k)$-bits syndrome associated with the error pattern with Hamming weight of $1$ ($\mathds{1}_i$). 

Shift registers are used in \cite{GRANDAB-VLSI} and \cite{ORBGRAND-TVLSI} to store syndrome of error patterns with a Hamming weight of $1$ ($\bm{s}_i$) as shown in Fig. \ref{fig:arch_1bit}. To test these error patterns, all of the rows of the shift register ($\bm{s}_i$) are combined with the syndrome of the received vector ($\bm{s}_c$) using a network of XOR gates. Following that, each of the $n$ syndromes obtained ($\bm{s}_i\oplus\bm{s}_c$) is NOR reduced and fed to a priority encoder, which chooses the test error pattern that meets the codebook membership criteria (\ref{eq:one-bit-constraint}). Each NOR-reduce output is $1$ if and only if all of the bits of the syndromes computed by  (\ref{eq:constraint}) are 0. 

Furthermore, the proposed GRANDAB \cite{GRANDAB-VLSI} and ORBGRAND \cite{ORBGRAND-TVLSI} decoders use the linearity property of the underlying code to combine $l$ syndromes of error patterns with a Hamming weight of $1$ ($\bm{s}_i$) to generate syndromes corresponding to an error pattern with a Hamming weight of $l$ ($\bm{s}_{1,2\ldots,l} = \bm{H}\cdot\mathds{1}_1^\top \oplus \bm{H}\cdot\mathds{1}_2^\top \ldots \oplus\bm{H}\cdot\mathds{1}_l^\top $). To understand the details of the VLSI implementation that is used to check error patterns with Hamming weight $\geq1$, we refer the reader to \cite{GRANDAB-VLSI} and \cite{ORBGRAND-TVLSI}. 

\subsection{VLSI Architecture for baseline ORBGRAND}

Figure \ref{fig:ORBarch} depicts the top-level ORBGRAND VLSI architecture \cite{ORBGRAND-TVLSI}, which can decode any linear block code with a length of $n$ and code rate $R\geq0.75$. The proposed architecture takes a vector of channel observation values ($\bm{y}$) as input and returns the estimated word $\hat{\bm{u}}$ as output. Any matrix can be loaded into $(n-k)\times n\text{-bit}$ \textit{H memory} at any time to support various codes and rates. The hard-demodulated vector $\hat{\bm{y}}$ is subjected to a syndrome check (\ref{eq:constraint}) in the first phase of decoding. If the syndrome ($\bm{s}_c$) is verified ($\bm{s}_c=\bm{0}$), decoding is presumed to be successful. Otherwise, the \textit{decoding core} generates the TEPs ($\bm{e}$) in the logistic weight order and applies them to $\hat{\bm{y}}$, after which the resulting vector $\hat{\bm{y}}\oplus\bm{e}$ is checked for codebook membership (\ref{eq:constraint}). If any of the tested syndrome combinations satisfy the parity check constraint (\ref{eq:constraint}), the 2D priority encoder is used in conjunction with the \textit{controller} module to forward the respective indices to the word generator module, where $P$ multiplexers are used to convert the sorted index values to their appropriate bit-flip locations.

\begin{figure}
  \centering
  \includegraphics[width=0.43\textwidth]{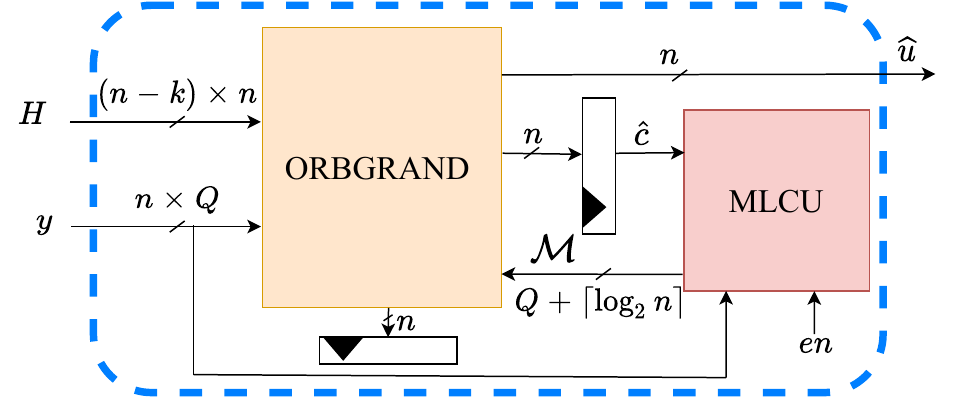}
  \caption{Proposed LGRAND VLSI Architecture}
  \label{fig:ListGRAND} 
  \vspace*{-1em}
\end{figure}

\begin{figure}
  \centering
  \includegraphics[width=0.55\textwidth]{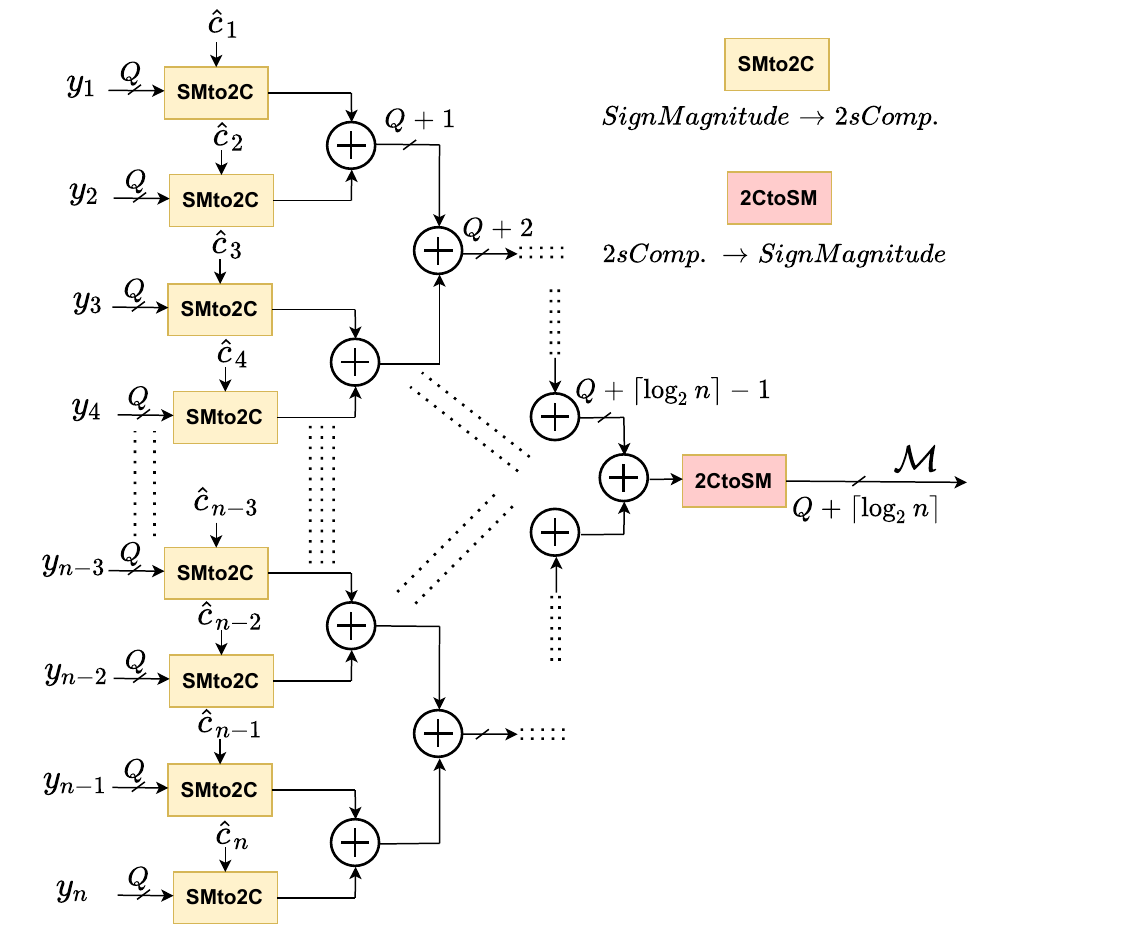}
  \vspace*{-1em}
  \caption{Maximum Likelihood Computation Unit ($\mathcal{M}\leftarrow \sum_{i=1}^{n}[(-1)^{\hat{\bm{c}}_i}\bm{y}_i$])}
  \label{fig:MLU} 
  \vspace*{-1em}
\end{figure}

\subsection{Proposed LGRAND VLSI Architecture}\label{sec:LGRAND-VLSI}

Fig.~\ref{fig:ListGRAND} depicts the proposed VLSI architecture for LGRAND, which builds up on the VLSI architecture for ORBGRAND \cite{ORBGRAND-TVLSI} and adds module \textit{Maximum Likelihood Computation Unit (MLCU)}. As described in section \ref{sec:List-GRAND}, the LGRAND selects the most likely codeword from a list ($\mathcal{L}$) of candidates. In the proposed LGRAND VLSI architecture, the ORBGRAND decoder works with the \textit{MLCU} to select the most likely codeword.

The proposed \textit{MLCU} unit's microarchitecture is depicted in Fig.~\ref{fig:MLU}. The \textit{MLCU} takes two inputs, $\hat{\bm{c}}$ and $n\times~Q$-bit $\bm{y}$, where $Q$ is the quantization width, and outputs the $Q+\left\lceil\log_2n\right\rceil$-bit value $\mathcal{M}$. An adder tree with $\log_2n$ stages is used to add the elements ($\bm{y}_i, \forall i \in [1, n]$) of the $\bm{y}$ vector to compute the likelihood $\mathcal{M}\leftarrow \sum_{i=1}^{n}[(-1)^{\hat{\bm{c}}_i}\bm{y}_i]$ \cite{ListRPA}. Furthermore, the components \textit{SMto2C} and \textit{2CtoSM} are used to convert from sign-magnitude to 2's complement form to facilitate signed addition and from 2's complement to sign-magnitude representation to facilitate comparison.

In the proposed LGRAND VLSI architecture shown in Figure ~\ref{fig:ListGRAND}, the ORBGRAND decoder delivers the estimated codeword ($\hat{\bm{c}}$) to the \textit{MLCU}, which computes the likelihood value $\mathcal{M}$ for the estimated codeword ($\hat{\bm{c}}$), then the \textit{MLCU} returns $\mathcal{M}$ to the ORBGRAND decoder. Please note that the list of estimated candidate codewords ($\mathcal{L}$) is not stored in a separate memory in the proposed LGRAND hardware; instead, as soon as an estimated codeword ($\hat{\bm{c}}$) is available at the ORBGRAND decoder, it is passed to the \textit{MLCU}.

The decoding core compares the likelihood values of the currently estimated codeword ($\mathcal{M}_{curr}$) and the previously estimated codeword ($\mathcal{M}_{prev}$). If and only if the $\mathcal{M}_{curr}$ value is higher than the $\mathcal{M}_{prev}$ value, the previous estimated codeword is replaced with the new one. If not, the decoding core retains the previous estimated codeword. The decoding core always maintains the estimated codeword with the highest likelihood by repeating this process for each successive estimated codeword. Finally, the original message ($\hat{\bm{u}}$) is retrieved from the estimated codeword with the highest likelihood ($\hat{\bm{c}}_{final} \leftarrow \argmax\limits_{\hat{\bm{c}}\in \mathcal{L}} \sum_{i=1}^{n}[(-1)^{\hat{\bm{c}}_i}\bm{y}_i]$\cite{ListRPA}) using $\bm{G}^{-1}$, and the decoding process is completed.

\begin{table}[!t]
\centering
\caption{\label{table:tableLGRAND}TSMC 65 nm CMOS Synthesis Comparison for LGRAND (LW$\leq$96, HW$\leq$8, $\delta\leq30$) with ORBGRAND (LW$\leq$96, HW$\leq$8) for $n=128/127$ and $0.75\leq~R~\leq1$.}
\begin{adjustbox}{max width=\columnwidth}
\begin{tabular}{lrrr}
\toprule
                             & & LGRAND      & ORBGRAND\cite{ORBGRAND-TVLSI}      \\
                              \cmidrule(l){3-3}\cmidrule(l){4-4}
Parameters                   & & {(LW$\leq$96, HW$\leq$8, $\delta\leq30$)} & {(LW$\leq$96, HW$\leq$8)} \\
Technology (nm)              & & {65}                 & 65                                    \\
Supply (V)                   & & {0.9}               & 0.9                                   \\
Max. Frequency (MHz)         & & {454}               & 454                                   \\
Area (mm\textsuperscript{2})  & & {2.38} & 2.27                       \\
W.C. Latency ($\mu$s)        &  & {205.76}     &  205.76                \\
Avg. Latency (ns)            & &2.2$^a$        & 2.2$^a$                    \\
\cmidrule(lr){2-4}
\multirow{5}{*}{{W.C. T/P (Mbps)}} & {$k=104^b$} & {0.505}              & {0.505}                                   \\
                              & {$k=105^c$} & {0.510}             & {0.510}                                   \\
                              & {$k=106^d$} & {0.515}              & {0.515}                                   \\
                              & {$k=112^e$} & {0.544}                  & {0.544}                                  \\
                              & {$k=113^f$} & {0.549}              & {0.549 }                                  \\
\cmidrule(lr){2-4}
\multirow{5}{*}{{Avg. T/P (Gbps)}} & {$k=104^b$} & {47.27}              & {47.27}                                   \\
                              & {$k=105^c$} & {47.72}              & {47.72}                                   \\
                              & {$k=106^d$} & {48.18}              & {48.18}                                   \\
                              & {$k=112^e$} & {50.90}                 & {50.90}                                  \\
                              & {$k=113^f$} & {51.36}              & {51.36}                                   \\
\cmidrule(lr){2-4}
{Power (mW)}           & &{{146.27}}       & {134.43}                     \\
\cmidrule(lr){2-4}
\multirow{5}{*}{{Energy per Bit (pJ/bit)}} & {$k=104^b$} & {3.09}              & {2.84 }                                  \\
                              & {$k=105^c$} & {3.06}              & {2.81}                                   \\
                              & {$k=106^d$} & {3.03}              & {2.79}                                   \\
                              & {$k=112^e$} & {2.87}  & {2.64}              \\
                              & {$k=113^f$} & {2.84}              & {2.62}                                   \\
\cmidrule(lr){2-4}
\multirow{5}{*}{{Area Efficiency (Gbps/mm\textsuperscript{2})}} & {$k=104^b$} & {19.86}              & {20.82}                                   \\
                              & {$k=105^c$} & {20.05}              & {21.02}                                   \\
                              & {$k=106^d$} & {20.24}              & {21.22}                                   \\
                              & {$k=112^e$} & {21.38}  & {22.42}              \\
                              & {$k=113^f$} & {21.58}              & {22.62}                                   \\
\cmidrule(lr){2-4}
Code compatible               & & {Yes}     & Yes                  \\
 Rate compatible              & & {Yes}     & Yes                \\ 
\bottomrule
\multicolumn{4}{l}{\footnotesize $^a$ For $\frac{E_b}{N_0}$ $\geq8.5$dB (Fig. \ref{fig:lat_tp}), $^b$ CRC Code (128,104), $^c$ Polar code (128,105+11) } \\
\multicolumn{4}{l}{\footnotesize $^d$ BCH Code (127,106), $^e$ CRC Code (128,112), $^f$ BCH Code (127,113) } \\
\multicolumn{4}{l}{\footnotesize $\text{Information Throughput (Gbps)}=\frac{{k}}{\text{Decoding Latency (ns)}}$} \\
\multicolumn{4}{l}{\footnotesize $\text{Energy per Bit (pJ/bit)}=\frac{\text{Power (mW)}}{\text{Avg. T/P (Gbps)}}$, $\text{Area Efficiency (Gbps/mm\textsuperscript{2})}=\frac{\text{Avg. T/P (Gbps)}}{\text{Area (mm\textsuperscript{2})}}$} \\
\end{tabular}
\end{adjustbox}
\end{table}


\begin{table}[!t]
\centering
\caption{\label{table:tableLGRANDPolar}{TSMC 65 nm CMOS Synthesis Comparison for LGRAND (LW$\leq$96, HW$\leq$8, $\delta\leq30$) with F.L. ORBGRAND \cite{carloORB} Decoder for 5G NR CA-polar code (128,105+11).}}
\begin{adjustbox}{max width=\columnwidth}
\begin{tabular}{@{}lrrr@{}}
\toprule
                             & {LGRAND}      & {F.L. ORBGRAND$^a$ \cite{carloORB}}      \\
                              \cmidrule(l){2-2}\cmidrule(l){3-3}
{Parameters}                   & (LW$\leq$96, HW$\leq$8, $\delta\leq30$) & ${Q_\text{max}}  = 2^{13}$ \\
{Technology (nm)}              & 65                & 7                                    \\
{Supply (V)}                   & 0.9               & 0.5                                  \\
{Max. Frequency (MHz)}         & 454               & 701                                   \\
{Area (mm\textsuperscript{2})}  & 2.38 & 3.70                       \\
{W.C. Latency (ns)}        &  205764.3     &  58.49                 \\
{Avg. Latency (ns)}            & 2.2$^b$        & 58.49                    \\
{W.C. T/P (Mbps) }              & 0.51         & 73610                                  \\
{Avg. T/P (Gbps)}               & 47.7         & 73.61                                   \\
{Power (mW) }                   & 146.27       & 170.84                    \\
{Energy per Bit (pJ/bit)}       & 3.06  & 2.32              \\
{Area Efficiency (Gbps/mm\textsuperscript{2})} & 20.04  & 19.89              \\
{Code compatible}               & Yes     & Yes                  \\
{ Rate compatible }             & Yes     & Yes                \\ 
\bottomrule
\multicolumn{3}{l}{\footnotesize $^a$ For $Q_\text{LUT} = 512$, $Q_\text{S} = 256$ , $T=34$} \\
\multicolumn{3}{l}{\footnotesize $^b$ For LGRAND with parameters $LW_\text{max}=96$, $HW_\text{max}=8$ and $\delta = 20$} \\
\end{tabular}
\end{adjustbox}
\end{table}

\begin{figure}[!t]
\centering
  \begin{tikzpicture}
    \begin{groupplot}[group style={group name=lat_tp, group size= 2 by 1, horizontal sep=10pt, vertical sep=10pt}, 
                      footnotesize,
                      height=.6\columnwidth,  width=.55\columnwidth,
                      xlabel=$\frac{E_{b}}{N_{0}}$ $(dB)$,
                      xmin=0, xmax=9, xtick={0,1,...,8},
                      ymode=log,
                      tick align=inside, 
                      grid=both, grid style={gray!30},
             ]

      \nextgroupplot[ylabel= Avg. Latency (cycles), ytick pos=left, y label style={at={(axis description cs:-0.15,.5)},anchor=south}, ymin=1, ymax = 2e4]
        \addplot[smooth, Paired-9, semithick]  table[x=SNR, y=LAT] {data/CRC/128_112/Lat_CRC_ORBGRAND_128_112.txt};\label{gp:plot1}
        \addplot[smooth,Paired-7, semithick]  table[x=SNR, y=LAT] {data/CRC/128_112/Lat_CRC_ListGRAND_128_112.txt};\label{gp:plot2}
        \addplot[smooth, Paired-5, semithick]  table[x=SNR, y=LAT] {data/BCH/127_113/Lat_BCH_ORBGRAND_127_113.txt};\label{gp:plot3}
        \addplot[smooth,Paired-3, semithick]  table[x=SNR, y=LAT] {data/BCH/127_113/Lat_BCH_ListGRAND_127_113.txt};\label{gp:plot4}

        \coordinate (top) at (rel axis cs:0,1);

      \nextgroupplot[ylabel=Avg. T/P (Mbps), ytick pos=right,y label style={at={(axis description cs:1.33,.5)},anchor=south}, ,ymax = 1e5]
        \addplot[smooth, Paired-9, semithick]  table[x=SNR, y=TP] {data/CRC/128_112/Lat_CRC_ORBGRAND_128_112.txt};
        \addplot[smooth, Paired-7, semithick]  table[x=SNR, y=TP] {data/CRC/128_112/Lat_CRC_ListGRAND_128_112.txt};
        \addplot[smooth, Paired-5, semithick]  table[x=SNR, y=TP] {data/BCH/127_113/Lat_BCH_ORBGRAND_127_113.txt};
        \addplot[smooth, Paired-3, semithick]  table[x=SNR, y=TP] {data/BCH/127_113/Lat_BCH_ListGRAND_127_113.txt};

        \coordinate (bot) at (rel axis cs:1,0);
    \end{groupplot}
    \node[below = 1cm of lat_tp c1r1.south] {(a) : Avg. Latency};
    \node[below = 1cm of lat_tp c2r1.south] {(b) : Avg. Info. Throughput};
    \path (top|-current bounding box.north) -- coordinate(legendpos) (bot|-current bounding box.north);
    \matrix[
        matrix of nodes,
        anchor=south,
        draw,
        inner sep=0.2em,
        draw
      ]at(legendpos)
      {
        \ref{gp:plot1}& \footnotesize CRC (128,112) ORBGRAND ($LW_\text{max}=96$, $HW_\text{max}=8$) \\
        \ref{gp:plot3}& \footnotesize BCH (127,113) ORBGRAND ($LW_\text{max}=96$, $HW_\text{max}=8$) \\
        \ref{gp:plot2}& \footnotesize CRC (128,112) LGRAND ($LW_\text{max}=96$, $HW_\text{max}=8$, $\delta = 24$) \\
        \ref{gp:plot4}& \footnotesize BCH (127,113) LGRAND ($LW_\text{max}=96$, $HW_\text{max}=8$, $\delta = 25$) \\
        };
  \end{tikzpicture}
  \vspace*{-2em}
  \caption{\label{fig:lat_tp}Comparison of average latency and average information throughput for the ORBGRAND\cite{ORBGRAND-TVLSI} VLSI architecture and the proposed LGRAND VLSI architecture for CRC Code (128,112) and BCH code (127,113).}
  \vspace*{-1em}
\end{figure}
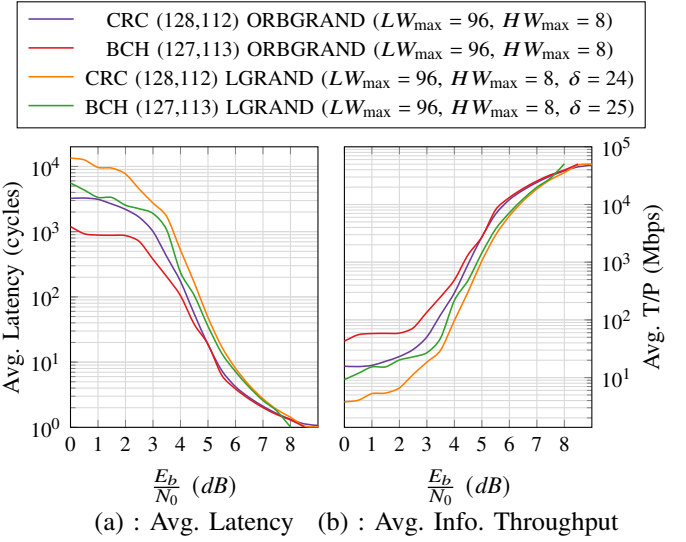
\subsection{Implementation results} \label{sec:Implementation}

The proposed LGRAND with parameters (LW$\leq$96, HW$\leq$8, $\delta\leq30$) has been implemented in Verilog HDL and synthesized with Synopsys Design Compiler using general-purpose TSMC 65 nm CMOS technology. Furthermore, the proposed LGRAND VLSI implementation is compared to the ORBGRAND\cite{ORBGRAND-TVLSI} VLSI implementation with parameters (LW$\leq$96, HW$\leq$8), and the synthesis results for $n = 128/127$ and code-rate $R$ ($0.75\leq~R~\leq1$) are shown in Table \ref{table:tableLGRAND}. Both designs, as shown in Table \ref{table:tableLGRAND}, are validated using test benches generated by the proposed hardware's bit-true C model. The input channel LLRs are quantized on 5 bits, with 1 sign bit and 3 bits for the fractional part. To ensure accuracy in power measurements, switching activities from real test vectors are extracted for both hardware architectures shown in Table \ref{table:tableLGRAND}.

The proposed LGRAND (LW$\leq$96, HW$\leq$8, $\delta\leq30$) has a $4.84\%$ area overhead over ORBGRAND (LW$\leq$96, HW$\leq$8), resulting in $\sim4.6\%$ less area efficiency than ORBGRAND. Furthermore, the proposed LGRAND is $7.7\%-8.2\%$ less energy efficient than the ORBGRAND (LW$\leq$96, HW$\leq$8). However, for decoding CRC code $(128,112)$, the proposed LGRAND with parameters ($LW_\text{max}=96$, $HW_\text{max}=8$, $\delta = 24$) outperforms ORBGRAND by $0.5$ dB at the target FER of $10^{-7}$, as illustrated in Fig. \ref{fig:grand_crc_128_112} (a). Similarly, as shown in Fig \ref{fig:grand_bch_2} (a), the proposed LGRAND with parameters $LW_{max}=96$, $HW_{max}=8$ and $\delta=25$ outperforms ORBGRAND by $0.75$ dB at a target FER of $10^{-7}$ for decoding BCH code (127,113).

The proposed LGRAND implementation can support a maximum frequency of $454~\text{MHz}$. One clock cycle corresponds to one time step because we do not consider any pipelining technique for the ORBGRAND decoder core. The proposed LGRAND architecture achieves a worst-case information throughput (W.C. T/P) of $0.5-0.549$-Mbps for different classes of channel codes as shown in Table \ref{table:tableLGRAND}. The average latency, on the other hand, is significantly smaller than the worst-case latency, especially at the higher $\frac{E_b}{N_0}$ region. The average latency is computed using the bit-true C model, of the proposed hardware, after taking into account at least 100 frames in error for each $\frac{E_b}{N_0}$ point. As channel conditions improve, the average latency for both ORBGRAND and LGRAND decreases until it reaches only 1 cycle per decoded codeword, as illustrated in Fig. \ref{fig:lat_tp}(a), resulting in an average latency of $2.2ns$ (corresponding to a maximum clock frequency of $454~\text{MHz}$). The average information throughput, which is the inverse of average latency, is depicted in Fig. \ref{fig:lat_tp} (b). It should be noted that the average information throughput increases with $\frac{E_b}{N_0}$, reaching values of $47.27-51.36$ Gbps.

The proposed LGRAND hardware is also compared to a recently proposed state-of-the-art Fixed-Latency (F.L) ORBGRAND decoder \cite{carloORB}, and the comparison results are presented in Table \ref{table:tableLGRANDPolar}. The F.L. ORBGRAND decoder deploys $T$ decoder pipeline stages and stores the test error patterns for the ORBGRAND decoding process in $T-2$ $Q_s\times{n}$-bit \textit{pattern memories}. When decoding 5G NR CRC-aided polar code (128,105+11), the F.L. ORBGRAND decoder can provide a maximum information throughput of $73.61$ Gbps with a fixed latency of $58.49ns$ owing to the highly pipelined VLSI architecture. The proposed LGRAND hardware, however, can achieve an average information throughput of $47.7$Gbps for the same polar code (128,105). At the target FER of $10^{-7}$, the proposed LGRAND with parameters ($LW_\text{max}=96$, $HW_\text{max}=8$, $\delta = 20$) outperforms both the ORBGRAND and F.L. ORBGRAND decoder \cite{carloORB} by $\sim0.5dB$ and can perform similarly to SGRAND, as depicted in Fig. \ref{fig:grand_polar}. Please note that scaling is not used to compare LGRAND and F.L. ORBGRAND \cite{carloORB} due to the vast disparity in the technology nodes employed (65nm vs 7 nm). The F.L. ORBGRAND \cite{carloORB} and LGRAND decoders are both code and rate compatible and can decode any code.

\section{Conclusion}
Soft GRAND (SGRAND) and Ordered Reliability Bits GRAND (ORBGRAND) are soft-input variants of GRAND, a universal decoder for short-length and high-rate codes. SGRAND delivers Maximum Likelihood (ML) decoding performance but is not suitable for parallel hardware implementation. ORBGRAND is suitable for parallel hardware implementation, however its decoding performance is inferior to SGRAND. In this paper, we introduced List-GRAND (LGRAND), a technique for improving the decoding performance of ORBGRAND. The proposed LGRAND includes parameters that can be tweaked to match the decoding performance and complexity budget of a target application. Furthermore, with the appropriate choice of parameters, LGRAND achieves decoding performance comparable to SGRAND.  Numerical simulation results show that the proposed LGRAND achieves a $0.5-0.75$dB performance gain over ORBGRAND for channel codes of different classes (BCH, CA-Polar, and CRC) at a target FER of $10^{-7}$. LGRAND, like ORBGRAND, lends itself to parallel hardware implementation. According to the VLSI implementation results, the proposed LGRAND has a $4.84\%$ area overhead over the ORGRAND hardware implementation. Furthermore, the proposed LGRAND VLSI architecture can achieve an average information throughput of $47.27-51.36$ Gbps for linear block codes of length $127/128$ and different code-rates.


%

\appendices

\ifCLASSOPTIONcaptionsoff
  \newpage
\fi



%



\bibliographystyle{IEEEbib}
\bibliography{refs}

\end{document}